\documentclass{emulateapj-rtx4}
\usepackage{natbib}
\usepackage{lscape}
\usepackage{hyperref}
\usepackage{morefloats}
\newcommand{\simless}{\mathbin{\lower 3pt\hbox
     {$\rlap{\raise 5pt\hbox{$\char'074$}}\mathchar"7218$}}}
\newcommand{\simgreat}{\mathbin{\lower 3pt\hbox
     {$\rlap{\raise 5pt\hbox{$\char'076$}}\mathchar"7218$}}}

\newcommand{\teff}{T$_{\rm eff}$}
\newcommand{\teffs}{T$_{\rm eff}$s}

\newcommand{\angstrom}{\textrm{\AA}}
\def\Tiny{ \font\Tinyfont = cmr10 at 6pt \relax  \Tinyfont}

\shortauthors{Pecaut et~al.}
\shorttitle{Young Stellar Colors and Temperatures}

\begin{document}
\title{Intrinsic Colors, Temperatures and Bolometric Corrections of Pre-main Sequence Stars}
\author{Mark J. Pecaut, Eric E. Mamajek}
\affil{University of Rochester, Department of Physics and Astronomy, Rochester, NY 14627-0171, USA}

\begin{abstract}
  We present an analysis of the intrinsic colors and temperatures of
  5-30~Myr old pre-main sequence (pre-MS) stars using the F0 through
  M9 type members of nearby, negligibly reddened groups: $\eta$~Cha
  cluster, TW~Hydra Association, $\beta$~Pic Moving Group, and
  Tucana-Horologium Association. To check the consistency of spectral
  types from the literature, we estimate new spectral types for 52
  nearby pre-MS stars with spectral types F3 through M4 using optical
  spectra taken with the SMARTS 1.5-m telescope.  Combining these new
  types with published spectral types, and photometry from the
  literature (Johnson-Cousins $BVI_C$, 2MASS $JHK_S$ and {\it WISE}
  $W1$, $W2$, $W3$, and $W4$), we derive a new empirical spectral
  type-color sequence for 5-30~Myr old pre-MS stars.  Colors for
  pre-MS stars match dwarf colors for some spectral types and colors,
  but for other spectral types and colors, deviations can exceed
  0.3~mag.  We estimate effective temperatures (\teff) and bolometric
  corrections (BCs) for our pre-MS star sample through comparing their
  photometry to synthetic photometry generated using the BT-Settl grid
  of model atmosphere spectra. We derive a new \teff\, and BC scale
  for pre-MS stars, which should be a more appropriate match for T
  Tauri stars than often-adopted dwarf star scales. While our new
  \teff\, scale for pre-MS stars is within $\simeq$100~K of dwarfs at
  a given spectral type for stars $<$G5, for G5 through K6, the pre-MS
  stars are $\sim$250~K cooler than their main sequence counterparts.
  Lastly, we present (1) a modern \teff, optical/IR color, and
  bolometric correction sequence for O9V-M9V MS stars based on an
  extensive literature survey, (2) a revised Q-method relation for
  dereddening UBV photometry of OB-type stars, and (3) introduce two
  candidate spectral standard stars as representatives of spectral
  types K8V and K9V.
\end{abstract}
\keywords{
  open clusters and associations: individual($\eta$~Cha cluster, TW~Hydra Association, $\beta$~Pic Moving Group, Tucana-Horologium Association);  ---
  stars: pre-main sequence ---
  stars: fundamental parameters (colors, temperatures)}
\maketitle

\section{Introduction and Background}
Knowledge of the stellar intrinsic color locus is an essential
ingredient in studying young stellar populations.  Recently-formed
stars are typically either distant, and thus outside of the ``Local
Bubble'' of low reddening in the solar vicinity, 
or they are still embedded in their natal molecular cloud.
Hence, we cannot assume negligible reddening and extinction for most
pre-main sequence (pre-MS) stars.  Interstellar reddening is
conventionally estimated using tabulated intrinsic colors of dwarf
field stars on the main sequence \citep[e.g.,][]{kenyon1995}.
However, this likely introduces systematic errors in the analysis
since the pre-MS stars are in a different evolutionary stage than the
main sequence calibrators and may not exhibit ``standard'' dwarf
colors.  Accurate H-R diagram placement depends on accurate extinction
and effective temperature (\teff) estimates.  If the extinction or
\teff\, is systematically in error because of systematics in the
intrinsic color and \teff\, tabulations as a function of spectral
type, this will obviously introduce systematic errors in the H-R
diagram placement and ages and masses inferred from comparison to 
evolutionary tracks.  For pre-MS stars, systematic errors in ages
may systematically shift the inferred timescales for
protoplanetary disk dissipation and giant planet formation
\citep[e.g.,][]{mamajek2009,bell2013}.  The H-R diagram presents an 
opportunity for stellar theoretical evolutionary models to make 
contact with observations, but if our H-R diagram placement is 
plagued with systematic errors, this makes testing evolutionary 
models impossible.  Thus it is imperative that the intrinsic color 
and \teff\, scale be accurately known and as free of systematic 
errors as possible.

Previous studies have noted that the intrinsic colors of young stars
differ from that of main sequence stars
\citep[e.g.,][]{gullbring1998,luhman1999b,bell2012}.  \cite{stauffer2003}
investigated Pleiades (age $\sim$125~Myr; \citealt{stauffer1998})
zero-age main sequence K-stars exhibiting bluer $B$--$V$ colors as a
function of spectral type than their counterparts in Praesepe (age
$\sim$750~Myr; \citealt{gaspar2009}), and concluded that the effect
was age-dependent.  Their study identified starspots and plages as the
most likely cause of the bluer $B$--$V$ colors and concluded that all
young K dwarfs will exhibit this effect.  \cite{dario2010} constructed
a young star intrinsic color sequence in their study of the
star-formation history of Orion Nebula Cluster by merging synthetic
colors interpolated to a 2~Myr isochronal surface gravity with
empirical colors from \cite{kenyon1995}. However, this implicitly
charges the color discrepancy solely to lower surface gravity.
Furthermore, synthetic near-infrared colors such as $J$--$H$ and
$H$--$K_S$ do not follow observed intrinsic color sequences for
M-dwarfs redder than $V$--$K_S\simgreat 4.0$ \citep[see
e.g.,][]{casagrande2008}, so we do not expect synthetic colors will
accurately predict the sequences of young stars \citep[though, see
also][]{scandariato2012}.  \cite{luhman2010,luhman2010e} compiled a
list of the IR photospheric colors for young K4 through
L0-type objects by fitting the blue envelope of the spectral
type-color sequence of young, nearby stars from Taurus, Chamaeleon~I,
the $\eta$~Cha cluster, the $\epsilon$~Cha cluster and the TW Hydra
Association (TWA).  The \cite{luhman2010} tabulation is empirically
derived and thus does not depend on synthetic colors.

Here we offer an alternative and expanded pre-MS intrinsic color
tabulation by including optical $BVI_C$ colors, including earlier
spectral types, and using the young stars' spectral energy
distributions to estimate effective temperatures and construct a
temperature and bolometric correction scale.  In this work we examine
spectral types F0 through M9.5, but our temperature scale only extends
to types as late as M5.  
In Section~\ref{sec:sample} we describe
our sample, and in Section~\ref{sec:data} we describe the spectroscopy
and photometry data used for our analysis.  In
Section~\ref{sec:analysis} we describe our spectroscopy, the
derivation of our pre-MS intrinsic colors, and the derivation of our
effective temperature and bolometric correction scale for pre-MS
stars.  Finally, in Section~\ref{sec:discussion} we compare our
temperature scale and angular diameter estimates to previous results
in the literature.

\section{Sample Selection}\label{sec:sample}

Our sample consists of members of young ($\simless$30~Myr), nearby
moving groups including the $\beta$~Pic moving group, TW~Hydra
Association (TWA), Tucana-Horologium moving group (Tuc-Hor) and the
$\eta$~Cha cluster.  The members of these groups are all predominantly
pre-main sequence (with the exception of a handful of
intermediate-mass A-type stars, which we omit) and thus will allow us
to study the observed color differences between main sequence stars
and pre-main sequence stars.  $\beta$~Pic, TWA and Tuc-Hor members are
less than 75~pc distant and thus lie within the Local Bubble, within
which objects are subject to negligible reddening (E($B$--$V$)$<$
0.002, using N$_H \simless 10^{19}$~cm$^{-2}$ inside the local bubble
from \citealt{cox1987} and N(H~I)/E($B$--$V$) = 4.8$\times
10^{21}$~cm$^{-2}$~mag$^{-1}$ from \citealt{savage1979}).  $\eta$~Cha
is slightly more distant ($\sim$95 pc) but also has A$_V \simeq 0$
\citep{mamajek1999,luhman2004}.  The negligible interstellar reddening
for these stars allows us to use their intrinsic colors to tabulate an intrinsic
color-spectral type relation for young stars in the widely used
Johnson-Cousins $BVI_C$, Two Micron All Sky Survey
\citep[2MASS;][]{skrutskie2006} $JHK_S$ photometric bands and the
Wide-Field Infrared Survey Explorer \citep[WISE;][]{wright2010} $W1$,
$W2$, $W3$ and $W4$ infrared bands at 3.4\,$\mu$m, 4.6\,$\mu$m,
12\,$\mu$m, and 22\,$\mu$m, respectively.

Our sample was assembled from group membership lists from 
\cite{mamajek1999}, \cite{luhman2004}, \cite{lyo2004},
\cite{song2004}, \cite{zuckerman2004}, \cite{scholz2005},
\cite{torres2006}, \cite{lepine2009}, \cite{kiss2011},
\cite{schlieder2010}, \cite{rice2010b}, \cite{zuckerman2011},
\cite{shkolnik2011}, \cite{rodriguez2011}, \cite{schlieder2012a} and
\cite{schneider2012b}.  Following the \cite{weinberger2012} and
\cite{mamajek2005} studies, we reject TWA~22 as a member of TWA based
on its discrepant space motion.  However, we retain it as a member of
$\beta$~Pic, following \cite{teixeira2009}.  In addition, based on the
study of \cite{mamajek2005} and parallax data from
\cite{weinberger2012}, stars TWA~14, TWA~15A, TWA~15B, TWA~17,
TWA~18, TWA~19A, TWA~19B, and TWA~24 are likely members of the Lower
Centaurus-Crux subgroup of the Scorpius-Centaurus OB association and
thus may be subject to non-negligible reddening, so we exclude them
from our sample.  We include TWA~9 as a member of TWA, though
\cite{weinberger2012} reject it.  We discuss our justification for
including it in Appendix~\ref{sec:twa9}.  Our sample includes 54
members of $\beta$~Pic with spectral types F0-M8, 34 members of TWA
with spectral types K3-M9.5, 45 members of Tuc-Hor with spectral types
F2-M2, and 15 members of $\eta$~Cha with spectral types K5-M5.75.

\section{Data}\label{sec:data}
\subsection{Spectroscopy}

Though the objects in our sample have published spectral types, they
are from a variety of sources and resolutions.  In order to check the
consistency of spectral types in the literature, we obtain new
spectral types using a grid of standards from
\cite{keenan1988,keenan1989}, \cite{kirkpatrick1991} and
\cite{henry2002}.  We acquired low-resolution blue
($\sim$3700\angstrom-5200\angstrom) and red
($\sim$5600\angstrom-6900\angstrom) optical spectra from the SMARTS
1.5m telescope in Cerro Tololo, Chile for 52 members of $\beta$~Pic,
TWA and $\eta$~Cha.  The stars chosen for spectroscopy were selected
based on (1) target brightness and (2) optimizing telescope time to
avoid interfering with higher priority programs.  The faintest targets
would require prohibitively large exposure times with the RC
spectrograph on the SMARTS 1.5m telescope to obtain useful S/N for
spectral classification.  This spectroscopic sample includes stars
down to m$_V\sim$14~mag, with spectral types F3-M4.  Observations were
made in queue mode with the RC spectrograph between February 2011 and
July 2011.  The blue spectra were taken with the ``26/Ia'' setup which
consists of a grating with groove density of 600 grooves~mm$^{-1}$,
blaze wavelength 4450\angstrom\, and no filter.  The red spectra were
taken with the ``47/Ib'' setup which consists of a grating with groove
density of 831 grooves~mm$^{-1}$, blaze wavelength 7100\angstrom, and
a GG495 filter.  Both used a slit with of 110.5$\mu$m.  The resolution
for the blue and red spectra are $\sim$4.3~\angstrom and
$\sim$3.1~\angstrom, respectively.  One comparison lamp exposure,
HeAr for blue spectra and Neon for red, was taken immediately before
three consecutive exposures of each target.  The data were reduced
using the SMARTS RC Spectrograph IDL pipeline of Fred Walter
\citep{walter2004}\footnote{\url{http://www.astro.sunysb.edu/fwalter/SMARTS/smarts\_15msched.html\#RCpipeline}}.
The three images are median combined, bias-trimmed, overscan-
and bias-subtracted and flat-fielded.  The spectrum is
wavelength-calibrated and, as a final step, we normalize the spectra
to the continuum with a low order spline in preparation for spectral
classification.

\subsection{Photometry}

After compiling the list of nearby $\simless$30~Myr old stars, we
assembled the most precise available photometry from the literature,
listed in Table~\ref{tbl:inputdata}.  All stars in our list have
counterparts in the 2MASS Point Source Catalog.  A few objects are
known binaries but are unresolved in the 2MASS catalog.  In these
cases, we retain the primary in our lists but do not include the
secondary since it would be of limited use without distinct
near-infrared photometry.  Tuc-Hor member TYC~7065-0879-1 (K0V;
\citealt{torres2006}) is a 1.8$\arcsec$ binary, resolved in Tycho-2
\citep{hog2000} but unresolved in 2MASS.  The 2MASS PSF photometry
differs significantly from the 2MASS aperture photometry (e.g.,
$H_{\rm PSF}-H_{\rm AP} = 0.356$~mag), presumably due to a poorly fit
PSF to the unresolved binary.  Thus for TYC~7065-0879-1 we adopt
unresolved $BVI_C$ optical photometry and the unresolved 2MASS
aperture photometry.  All other objects in our sample have 2MASS PSF
photometry which agrees well with the aperture photometry (when
available) and therefore we simply adopt the PSF photometry.  We adopt
{\it WISE} bands $W1$, $W2$, $W3$, and $W4$ photometry from the {\it
  WISE} All-Sky Point Source Catalog, centered at 3.4, 4.6, 12, and 22
$\mu$m, respectively \citep{wright2010}.  Objects saturated in $W2$
($\simless$ 6.3~mag) exhibit a flux over-estimation
bias\footnote{\url{http://wise2.ipac.caltech.edu/docs/release/allsky/expsup/sec6_3c.html}},
so to avoid these biases we exclude $W2$ photometry with $W2 <
6.0$~mag.  For stars with {\it Hipparcos} catalog entries, we adopt
$V$ and $B$--$V$ photometry from that catalog \cite{esa1997}.  We then
fill missing $B$--$V$ photometry using Tycho-2 photometry ($B_T$,
$V_T$) converted to Johnson $B$--$V$ with the conversions of
\cite{mamajek2002,mamajek2002e}, resorting to the conversions in
\cite{hog2000} when $B_T-V_T > 2.0$. We adopted AAVSO Photometric
All-Sky Survey (APASS) Data Release 6 \citep{henden2012} $BV$ and SACY
\citep{torres2006} $BVI_C$ photometry where available.  Conservative
estimates for SACY $BVI_C$ photometric uncertainties obtained with the
FOTRAP instrument are 0.01~mag for stars brighter than $V\sim12$~mag
(C.A.O. Torres, 2012 private communication).  We only adopted $B$--$V$
colors when $\sigma_{B-V} < 0.08$~mag.  We adopted $V$--$I_C$
photometry from \cite{torres2006}, \cite{lawson2001} and the {\it
  Hipparcos} catalog, when it was directly observed (value ``A'' in
field H42), since a significant portion of the tabulated $V$--$I_C$
photometry in the {\it Hipparcos} catalog is inferred from photometry
in other bands or from the spectral type of the star.  Though it was
available for many of our objects, we did not adopt DEep Near-Infrared
Survey of the Southern Sky (DENIS) $i$ band photometry since it
saturates at $\sim$10.3~mag \citep{epchtein1997} and therefore most of
our objects are too bright to have reliable DENIS $i$ photometry.

\section{Analysis}\label{sec:analysis}
\subsection{Spectral Classification}

The optical spectra were visually classified by directly comparing
them with spectral standards using a custom spectral software tool,
{\it sptool\footnote{See
    \url{http://www.pas.rochester.edu/\~mpecaut/sptool} or
    \url{rumtph.org/pecaut/sptool/}.}}, described in
\citet{pecaut2012}.  F- and G-type standards are taken from Table~2 of
\citealt{pecaut2012}; K- and M-type standards are listed in
Table~\ref{tbl:spstands}.  For the blue spectra, the F-type stars were
classified using the strength and profile of the Balmer lines, with
particular attention to the wings of the lines in case the line depths
were filled in by chromospheric emission.  In addition, we use the
G-band at $\sim$4310\angstrom\, as it is a very useful temperature
indicator for solar metallicity F-type stars \citep{gray2009}.  For
G-type stars we use the G-band, Fe~I lines at 4046\angstrom,
4325\angstrom, and 4383\angstrom, the Ca~I line at 4227\angstrom, and
the Mg~Ib triplet at $\sim$5170\angstrom.  Spectral classification
using the features described here is discussed in greater detail in
\cite{gray2009}.

\begin{figure}
\begin{center}
\includegraphics[scale=0.45]{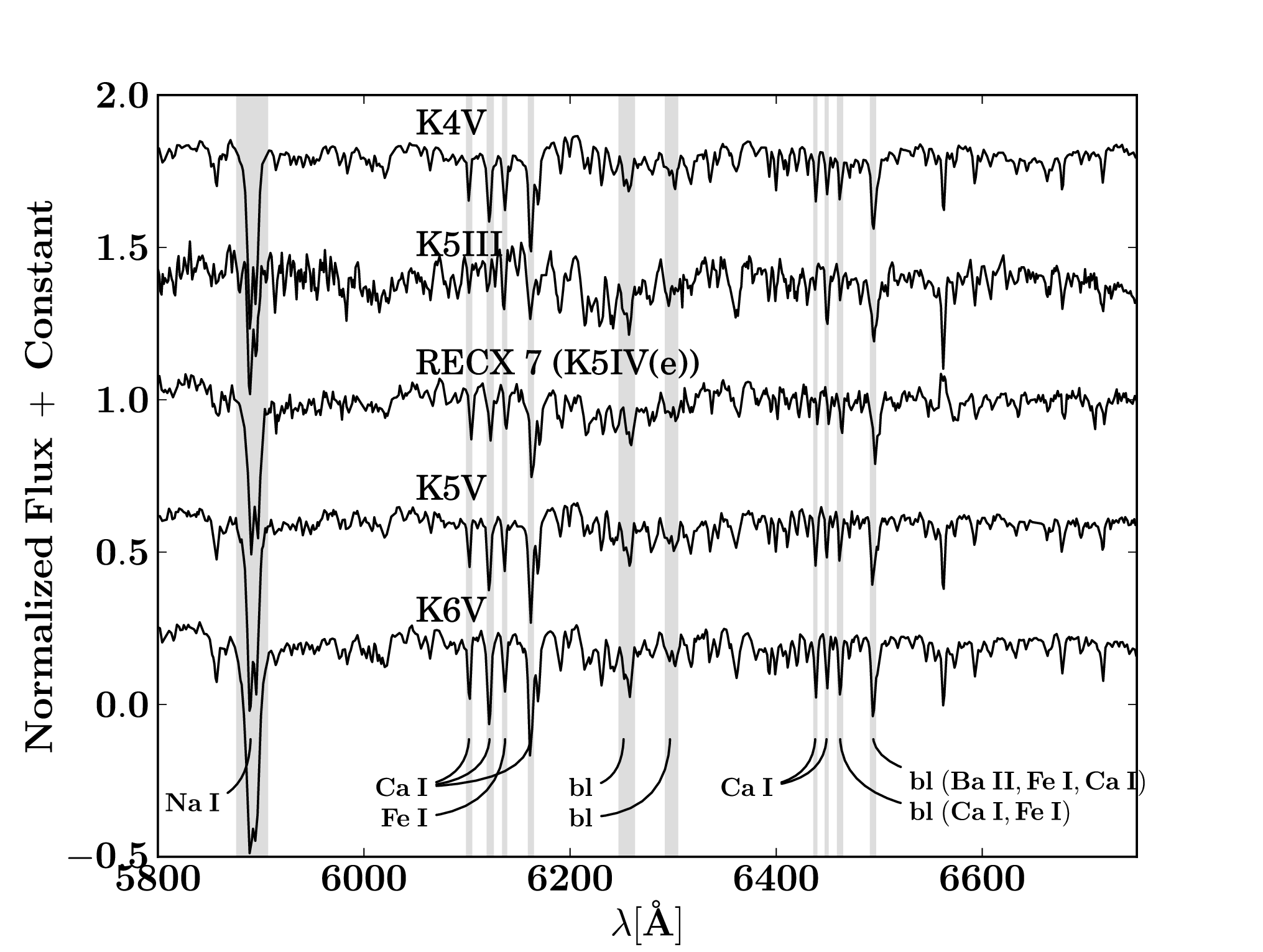}
\caption{\label{fig:sptclass}
    A spectrum of $\eta$~Cha member RECX~7 (K5IV(e)) with spectral
    standards K4Ve (TW~PsA), K5III (HD~82668), K5V (HD~36003),
    and K6Va (GJ~529).  The primary regions used for 
    spectral classification of K-type stars are highlighted in grey.
}
\end{center}
\end{figure}

For stars with red spectra ($\sim$5600\angstrom-6900\angstrom) only,
we first determine if it is a K- or M-type star based on the overall
appearance of the spectrum.  For K-type stars we obtain accurate
spectral classifications using the Ca~I lines at 6102\angstrom,
6122\angstrom, 6162\angstrom, and 6169\angstrom, the Fe~I lines at
6137\angstrom, the relative strength of the V~I and Fe~I blend at
6252\angstrom\, to Ti~I at 6258\angstrom, and the relative strength of
the V, Ti, and Fe blend at 6297\angstrom\, to the Fe~I blend at
6302\angstrom\footnote{ Many of these lines were identified using the
  VALD service \citep{kupka1999}.
  \url{http://vald.astro.univie.ac.at/}}.  We also made use of Ca~I
lines at $\lambda\lambda$\,6438 and 6449, the Ca~I/Fe~I blend at
6462\angstrom, the Fe~I, Ti~I and Cr~I blend at 6362\angstrom, the
Ba~II, Fe~I and Ca~I blend at 6497\angstrom, and for the latest
K-types, the TiO bands from $\sim$6651\angstrom-6852\angstrom.  For
M-type stars we use the Ca~I lines at 6122\angstrom\, and
6162\angstrom, but predominantly rely on TiO bands from
$\sim$5847\angstrom-6858\angstrom, $\sim$6080\angstrom-6390\angstrom,
and $\sim$6651\angstrom-6852\angstrom.  Following
\cite{gray2003,gray2006}, we assign spectral types K8 and K9, where
appropriate.  This is discussed in more detail in
Appendix~\ref{sec:k7_m0}.  Example spectra with the lines used are
shown in Figure~\ref{fig:sptclass}.

While obtaining temperature types for our sample we ignored the Na~I
doublet at $\sim5889/5896$\angstrom\, because it is sensitive to both
temperature and surface gravity and is thus useful in discriminating
between stars with dwarf-like, subgiant-like or giant-like gravity.
The young stars in our sample are pre-main sequence and thus may have
a Na~I doublet line similar to subgiants.  Once a temperature type had
been established, we compared the Na~I doublet to that of a dwarf and
a giant of the same temperature subclass, assigning the luminosity
class ``IV'' if the strength was intermediate between the dwarf and
giant, ``IV-V'' if the strength was very similar to that of a dwarf
but only slightly weaker, and ``V'' if the Na doublet was
indistinguishable from a dwarf.  The results of our spectral
classification are listed in Table~\ref{tbl:sptypes}.

\subsection{Synthetic Colors}

In order to compare observed colors to model atmosphere predictions
for the color locus and the predicted effects of surface gravity, we
compare our observed colors with synthetic colors calculated from the
``BT-Settl'' models from the Phoenix/NextGen group
\citep{hauschildt1999,allard2012} and the ``ATLAS9'' models from
\cite{castelli2004}.  The BT-Settl models offer synthetic spectra with
$400~K < $\teff$ < 70000~K$, $-0.5 < \log(g) < 5.5$ and $-4.0 < [M/H]
< +0.5$, with $\alpha$-element enhancement between +0.0 and +0.6~dex.
The ATLAS9 models offer synthetic spectra with $3500~K < $\teff$ <
50000~K$, $0.0 < \log(g) < 5.0$, $-5.5 < [M/H] < +0.5$ with
$\alpha$-element enhancement between +0.0 and +0.4~dex.  However,
since our objects are young and are in the solar neighborhood, we
assume solar metallicity with no $\alpha$-element enhancement.  This
is consistent with the findings of \cite{vianaalmeida2009}, who have
spectroscopically analyzed a small sample of these young stars,
obtaining $<$[Fe/H]$>$=-0.06$\pm$0.09~dex for a sample of nine Tuc-Hor
stars and [Fe/H]=-0.13$\pm$0.08~dex for $\beta$~Pic member HD~322990.
We computed synthetic colors, listed in Table~\ref{tbl:syncolors}, for
solar metallicity models with with $3.0 < \log(g) < 5.0$, $1400~K <
$\teff$ < 50000~K$ for the BT-Settl models and $3500~K < $\teff$ <
50000~K$ for the ATLAS9 models, with no $\alpha$-element enhancement.
Pre-MS stars have lower surface gravities than main sequence stars at
the same \teff\, but both should have $3.0 < \log(g) < 5.0$.  We wish
to evaluate model predictions of color trends as a function of surface
gravity, so we plot synthetic colors for both $\log(g)=3.0$ and 5.0.
A coeval population will have a surface gravity
which varies as a function of mass, so we also plot a sequence with
surface gravities given by a 20~Myr isochrone from the
\cite{baraffe1998} models.  We plot commonly used colors against
$V$--$K_S$.  We chose $V$--$K_S$ because it is available for nearly
all our objects, and it offers a very large baseline compared to other
colors so it is useful as a proxy for \teff.  To compute the synthetic
photometry for the models, we use the updated $BVI_C$ normalized
photonic bandpasses and zero points from \cite{bessell2012}, including
the additional zeropoints listed in their Table~5.  To compute the
2MASS $JHK_S$ synthetic photometry, we use the relative system
response (RSR) curves available on the IPAC
website\footnote{\url{http://www.ipac.caltech.edu/2mass/releases/allsky/doc/sec6_4a.html}}
with the zero magnitude flux given in \cite{rieke2008}.  Similarly,
for the WISE bands we use RSR curves available on the IPAC
website\footnote{\url{http://wise2.ipac.caltech.edu/docs/release/prelim/expsup/sec4_3g.html}}
with the zero magnitude flux given in \cite{jarrett2011}.  The ATLAS9
models are sparsely sampled past $\sim$10$\mu$m, with only 9 points
representing the flux density from 10$\mu$m to 160$\mu$m, so we linearly 
interpolate $\lambda^{4}F_\lambda$ from 10$\mu$m to 160$\mu$m and divide 
by $\lambda^4$ before performing synthetic photometry.
This is not necessary for the BT-Settl models because they are sampled
at 0.2\angstrom spectral resolution for $\lambda > 5.2 \mu$m.  The
BT-Settl models shown adopt the \cite{asplund2009} solar abundances while
the ATLAS9 models shown use the \cite{grevesse1998} solar abundances.
The computed synthetic colors are listed in Table~\ref{tbl:syncolors}.

\subsection{Empirical Colors of Dwarfs Versus Pre-MS Stars}\label{sec:color_color}

\begin{figure*}[]
\begin{center}
\includegraphics[scale=0.45]{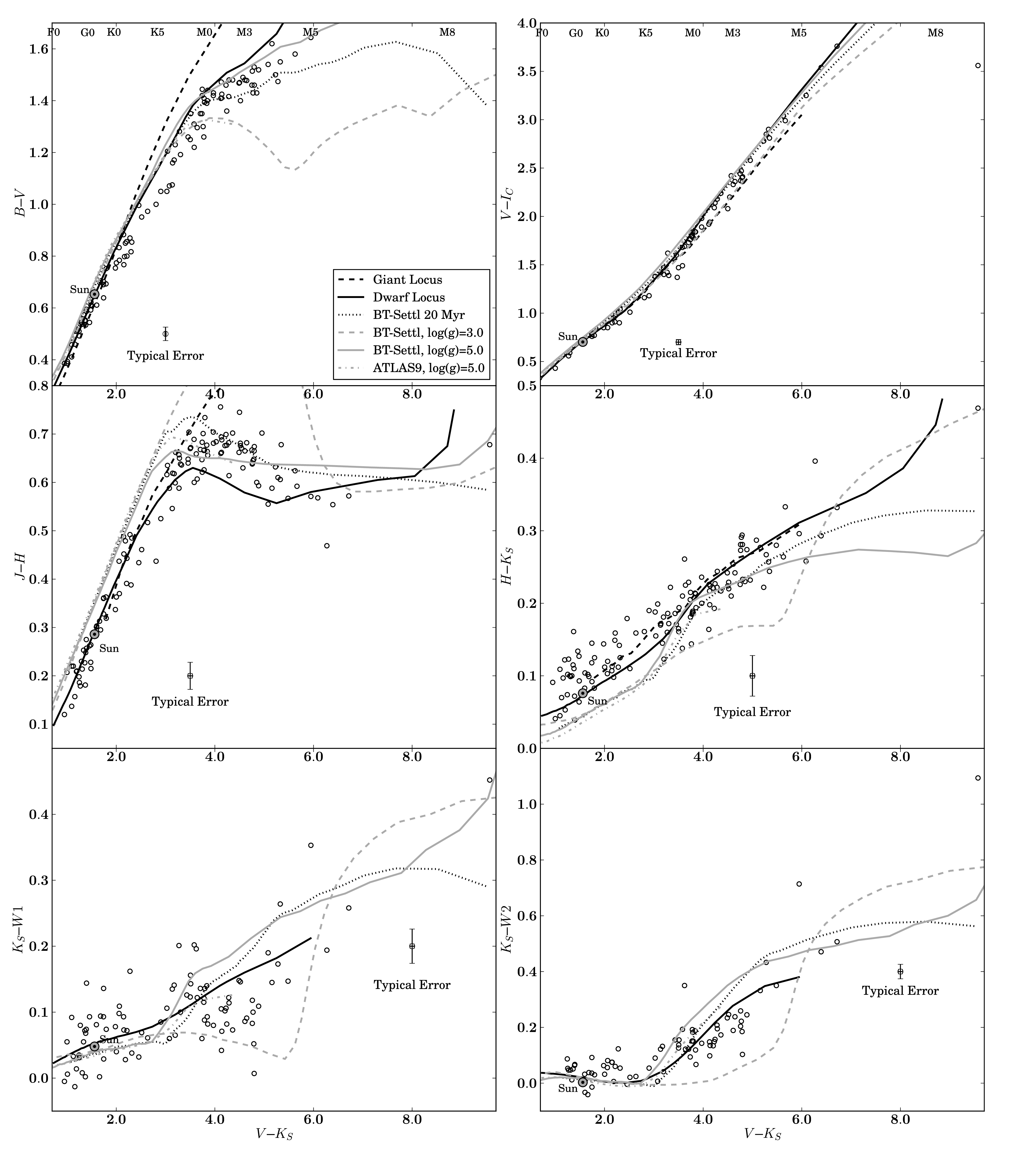}
\caption{\label{fig:color_color}
	Comparison of $B$--$V$, $V$--$I_C$, $J$--$H$, $H$--$K_S$, $K_S$--$W1$,
    and $K_S$--$W2$ versus $V$--$K_S$ of young stars from $\beta$~Pic, 
    $\eta$~Cha, TWA and Tuc-Hor moving groups (circles) with the dwarf color 
    locus described in Appendix~\ref{sec:dwarfcolors} and the giant color locus 
    from \cite{bessell1988}, except the $B$--$V$ giant locus, which is from 
    \cite{alonso1999}.  Spectral types corresponding to the $V$--$K_S$ colors 
    of dwarfs are plotted along the top.  Objects with a known near-IR or IR 
    excess have been excluded (see Table~\ref{tbl:inputdata}).
}
\end{center}
\end{figure*}

\begin{figure}
\begin{center}
\includegraphics[scale=0.45]{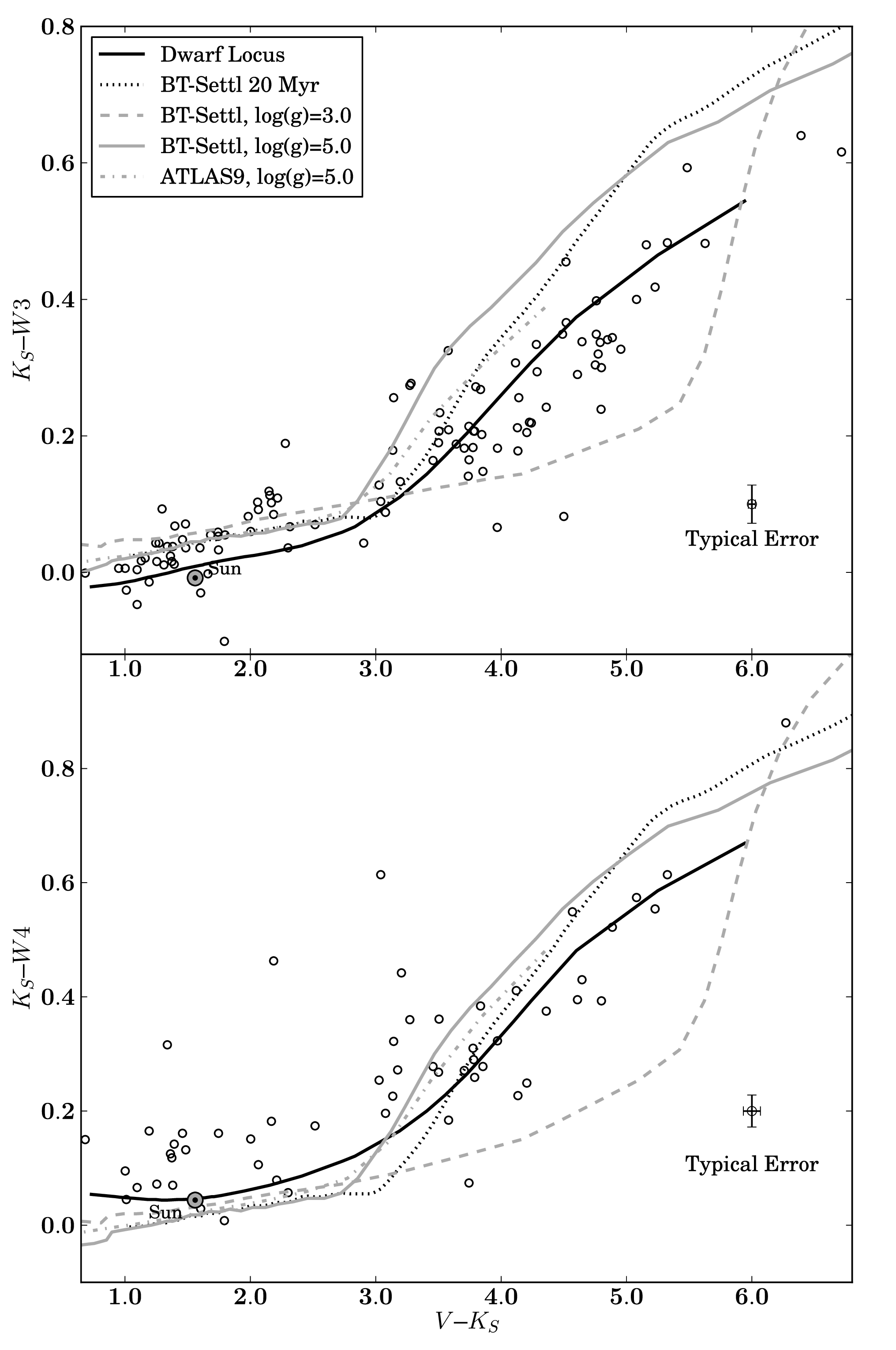}
\caption{\label{fig:color_color2}
Same as Figure~\ref{fig:color_color}, except $V$--$K_S$ versus $K_S$--$W3$ and
    $K_S$--$W4$.  
}
\end{center}
\end{figure}

To compare dwarfs colors with pre-MS colors, we plot color-color
diagrams for the young stars listed in
Table~\ref{tbl:inputdata}.  Figures~\ref{fig:color_color} and
\ref{fig:color_color2} show $V$--$K_S$ versus $B$--$V$, $V$--$I_C$,
$J$--$H$, $H$--$K_S$, $K_S$--$W1$, $K_S$--$W2$, $K_S$--$W3$ and
$K_S$--$W4$ for the young stars along with the dwarf sequence
described in Appendix~\ref{sec:dwarfcolors} (listed in
Table~\ref{tbl:ic_teff}) and the empirical giant sequence for $B$--$V$
from \cite{alonso1999} and for $V$--$I_C$, $J$--$H$, and $H$--$K_S$
from \cite{bessell1988} converted to the 2MASS photometric system with
the conversions of \cite{carpenter2001}.  For reference we include the
$BVI_C$ solar colors estimated by \cite{ramirez2012} and 2MASS $JHK_S$
and WISE $W1\,W2\,W3\,W4$ solar colors estimated by
\cite{casagrande2012}.

Color-color plots $V$--$K_S$ versus $B$--$V$ and $V$--$K_S$ versus
$J$--$H$ show the largest color difference between our young stars and
the dwarf locus.  Redward of $V$--$K_S \sim 2.0$~mag, young stars are
bluer in $B$--$V$ than the dwarf locus, and for $V$--$K_S \geq 4.0$
they are well-matched by the 20~Myr isochronal colors.  Models predict
the $B$--$V$ colors are bluer at lower surface gravity at a given
$V$--$K_S$, consistent with our observations, though the agreement is
not perfect.  Models predict little sensitivity to surface gravity for
$V$--$K_S$ versus $V$--$I_C$, consistent with the location of the
dwarf and giant locus as well as the placement of the young stars.
For $V$--$K_S$ versus $J$--$H$ locus, a bifurcation between the dwarf
and giant empirical locus occurs at $V$--$K_S$ $\sim$3~mag, which
corresponds to spectral type $\sim$K5.  This color split has been explained
by the models as an effect of surface gravity, due to the CO and
H$_2$O bands and H$^-$ opacity \citep{jorgensen1996}.  The young stars
in our sample have surface gravities intermediate between that of the
giants and dwarfs, and as a result they populate the region between
the the dwarf and giant loci.  For $V$--$K_S \leq 3.5$, the young
stars lie above the dwarf locus for colors $H$--$K_S$ and $K_S$--$W1$,
indicating that these two colors are redder for young stars at a given
$V$--$K_S$.  We exclude photometry for objects which have previously
identified infrared excesses in that respective infrared band, likely
due to a dusty circumstellar disk.  Excluded photometry is indicated in 
Table~\ref{tbl:inputdata}.

\subsection{Spectral Type-Color Sequence}

To define the intrinsic color sequence empirically, with the
constraint of satisfying the color-color plots, we first fit a spline
to spectral type versus $V$--$K_S$ and spectral type versus
$V$--$I_C$.  We then verify that these relations provide a good fit to
the $V$--$K_S$ versus $V$--$I_C$ color-color relation as well.  We
then fit splines to $V$--$K_S$ versus $J$--$H$ and $V$--$K_S$ versus
$H$--$K_S$ and use our spectral type-$V$--$K_S$ relation to anchor
$J$--$H$ and $H$--$K_S$ to spectral type.  Finally, we fit splines to
spectral type versus color for the colors $B$--$V$, $K_S$--$W1$,
$K_S$--$W2$, $K_S$--$W3$ and $K_S$--$W4$.  $V$--$I_C$ data is sparse
for types earlier than G5, but appears consistent with the dwarf
sequence, so we simply adopt the dwarf $V$--$I_C$ sequence for
spectral types F0 through G5 discussed in Appendix~\ref{sec:dwarfcolors}.  
In Figure~\ref{fig:spt_color} we see
that pre-MS stars later than K3 become {\it bluer} in $B$--$V$ than
their main sequence counterparts, while those hotter than K2 are
nearly indistinguishable from main sequence stars.
Figure~\ref{fig:spt_color} also shows that young stars G5 and later
have redder $V$--$K_S$ and $J$--$H$ colors than field dwarfs, while
those earlier than G5 have $V$--$K_S$ and $J$--$H$ colors
indistinguishable from field dwarfs.  Pre-MS stars have $H$--$K_S$
colors redder than field dwarfs between spectral types F0 and M2,
shown in Figure~\ref{fig:spt_color}.  The spectral type sequence for
$K_S$--$W1$, $K_S$--$W2$, $K_S$--$W3$ and $K_S$--$W4$
(Figures~\ref{fig:spt_color} and \ref{fig:spt_color2}) show larger
scatter than for the previously discussed colors, and greater care
must be taken to exclude those stars with a color excess due to the
presence of a circumstellar disk.  We have excluded photometry for
objects with infrared excesses flagged in 
Table~\ref{tbl:inputdata}.  AG~Tri was discussed in
\cite{rebull2008} as having a possible MIPS 24$\mu$m excess. We find
that it has a $K_S$--$W4$ color excess 4.5$\sigma$ above the young
color sequence.  We also identify HD~160305 and CD-54~7336 as having a
$K_S$--$W4$ color excess at 2.9$\sigma$ and 5.4$\sigma$ above the
young color sequence, so we also exclude them from the $K_S$--$W4$
fit.  Our pre-MS intrinsic color sequence is listed in
Table~\ref{tbl:young_colors}.

For some spectral type and color combinations, extinction estimates using these 
intrinsic colors will give different results than those which adopt dwarf colors.
For example, a typical unreddened pre-MS K0 star has a $V$--$K_S$ color 0.24~mag 
redder than a main-sequence K0.  If one estimated A$_V$ based on the stars $E(V-K_S)$
calculated using dwarf colors, then this star would appear to have 
$A_V$=1.12$E(V-K_S)\simeq$0.27~mag of artificial extinction, based on the 
apparent $V$--$K_S$ color difference between pre-MS and a main-sequence K0 stars 
(assuming a standard $R_V$=3.1 reddening law).  A 0.3~mag
systematic shift in H-R diagram placement would cause a 15~Myr old
K-type star to erroneously appear 10~Myr old.

\begin{figure*}[]
\begin{center}
\includegraphics[scale=0.45]{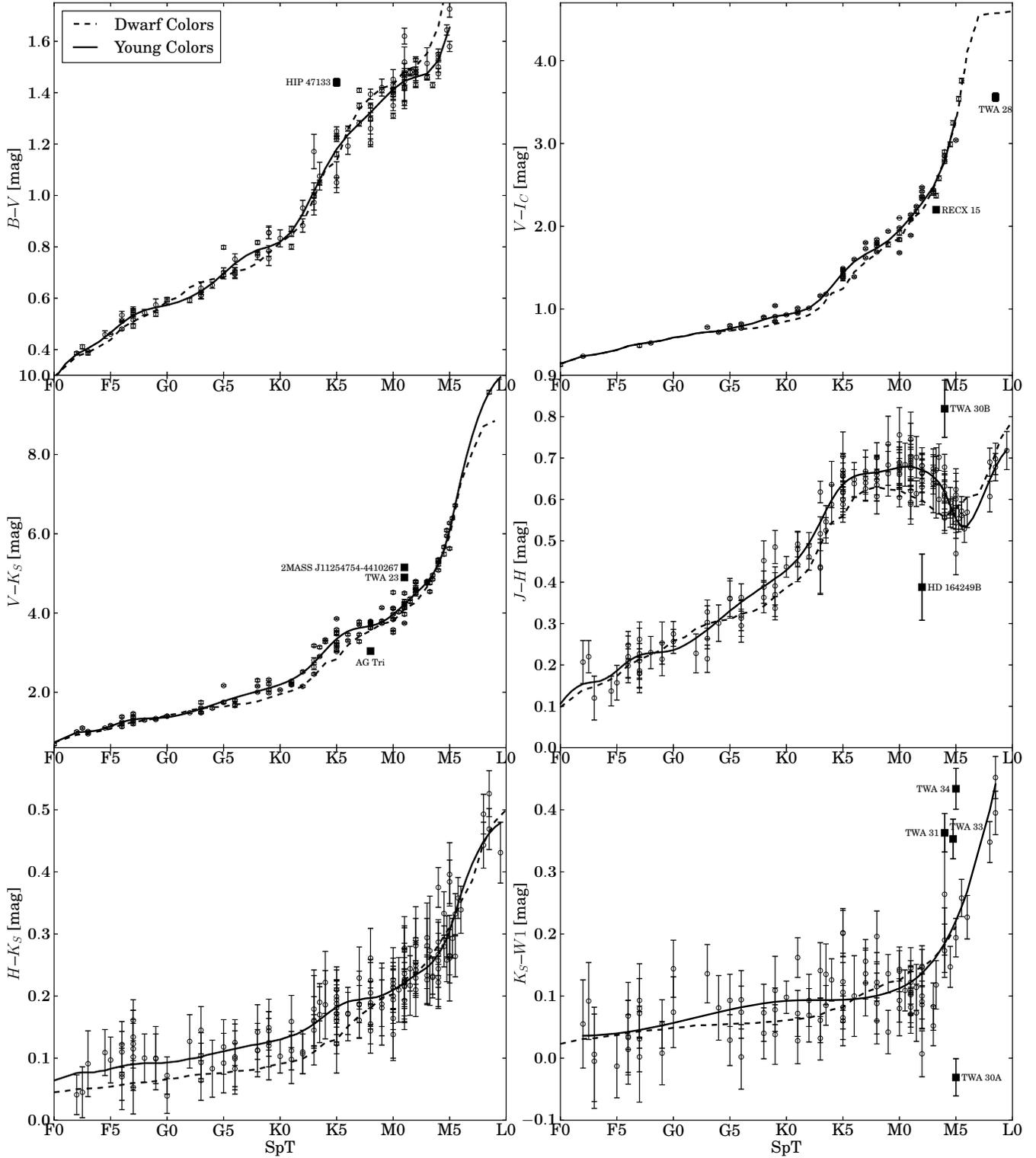}
\caption{\label{fig:spt_color}
    Comparison of $B$--$V$, $V$--$I_C$, $V$--$K_S$, $J$--$H$, $H$--$K_S$, and
    $K_S$--$W1$ of young stars from $\beta$~Pic, $\eta$~Cha, TWA
    and Tuc-Hor moving groups (circles) with the dwarf color sequence described
    in this work (dashed line).   The outliers (filled squares) were excluded 
    from the fit.
}
\end{center}
\end{figure*}

\begin{figure}
\begin{center}
\includegraphics[scale=0.45]{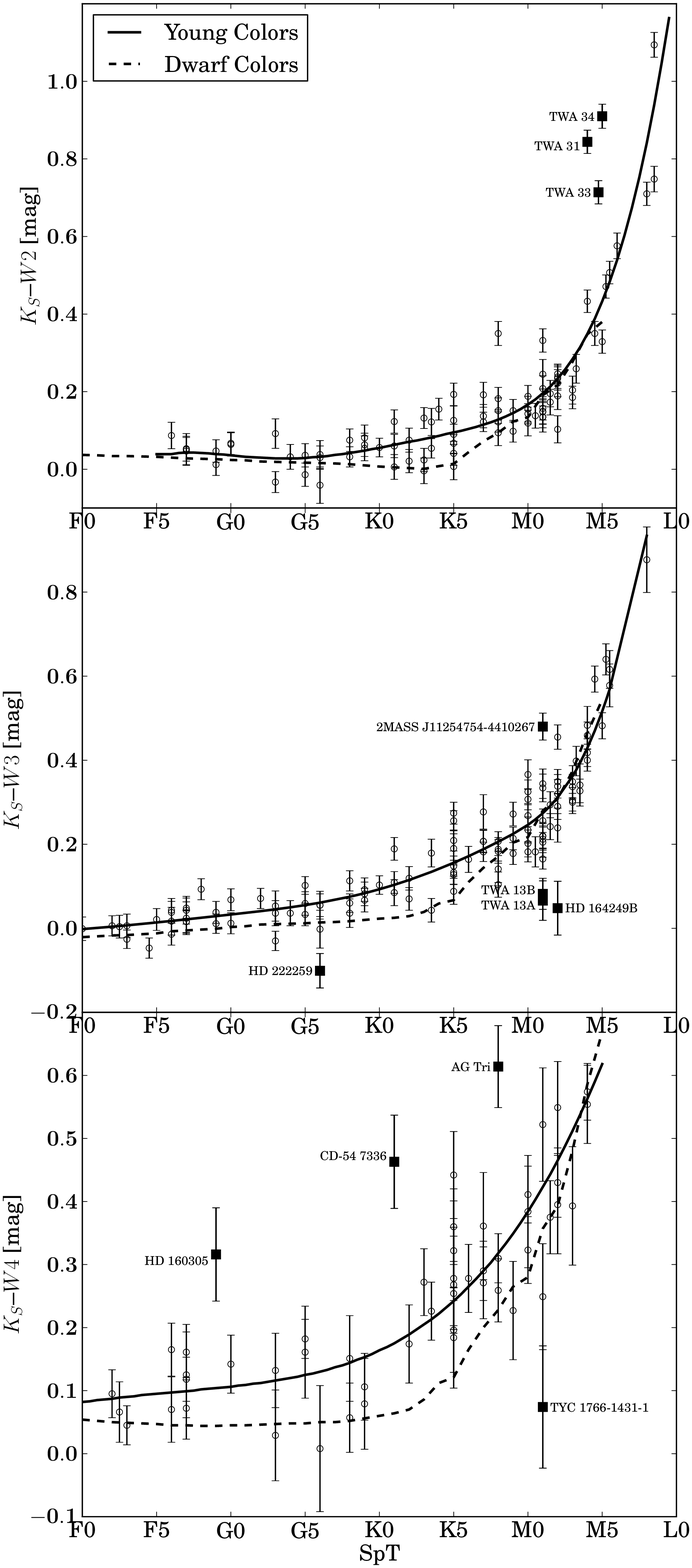}
\caption{\label{fig:spt_color2}
    Same as in Figure~\ref{fig:spt_color}, except showing colors $K_S$--$W2$,
    $K_S$--$W3$, and $K_S$--$W4$.  
    Outliers (filled squares) have been excluded from the fit, and 
    objects with known infrared excesses are not shown.
}
\end{center}
\end{figure}

\subsection{Temperature Scale}
\subsubsection{Technique}

The effective temperature (\teff) scale for giants
\citep[e.g.,][]{vanbelle1999} as a function of spectral type is
$\sim$700-400~K cooler than dwarfs for spectral types G8 through K5,
whereas M0 through M9 giants are $\sim$100-400~K hotter than dwarfs.
Since pre-MS stars have surface gravities intermediate
between dwarfs and giants, we expect that a pre-MS \teff\, scale will
be intermediate between dwarfs and giants \citep[e.g.,
][]{luhman2003}.

All \teff\, scales depend on models (e.g., atmospheric models, 
limb-darkening models) to some degree.  Arguably, the least 
model-dependent methods are those {\it direct} methods based on the angular 
diameter of the star, measured interferometrically or by lunar occultation 
methods.  While some of the stars in our sample are candidates for angular 
diameter measurements \citep[see][]{mccarthy2012}, only two have actual 
measurements in the literature (HR~9 and 51~Eri; \citealt{simon2011}; see 
Section~\ref{sec:discussion} for details).  There are also {\it indirect} 
methods, such as the infrared flux method (IRFM), performed by
\cite{alonso1999}, and more recently for M-dwarfs by \cite{casagrande2008}, 
or directly fitting spectral energy distributions to synthetic model 
photometry, as described by \cite{masana2006}.

Spectroscopically, young stars have been shown to exhibit more than one 
photospheric \teff \citep{gullbring1998,stauffer2003}, so fitting synthetic spectra 
to observed spectra will yield a different \teff\, depending on the spectral region 
selected for fitting.  An example of this is TW~Hydra, which has been consistently
typed as a late K star based on optical spectra (K7e, \citealt{delareza1989}; 
K6e, \citealt{hoff1998}; K6Ve, \citealt{torres2006}; K8IVe, this work) but near-IR
spectroscopy indicate a spectral type of M2.5V \citep{vacca2011}.  
We need a method to infer temperatures that will simultaneously take into account 
the observed optical-IR photometry.  Therefore we attempt to infer the effective 
temperatures by simultaneously fitting the observed photometry to synthetic models 
(the ``Spectral Energy Distribution Fitting'' (SEDF) method, see \citealt{masana2006}).  
The downside of this method is that we are using models which do not completely 
correctly predict the colors of young stars.  However, since the \teff\, is {\it defined} 
by the integrated spectral energy distribution (SED) and the stellar radius, the 
observed photometry is the most direct link to the effective temperature of objects 
in our sample.  We closely follow the formalism and methods of 
\cite{masana2006} and fit the observed photometry to models by minimizing $\chi^2$, 
defined as

\begin{eqnarray*}
  \chi^2 & = & \sum\limits_{i} \left(\frac{m_i - m_{i,{\rm syn}} - \mathcal{A}}{\sigma_{m_i}}\right)^2  \\
\end{eqnarray*}

With m$_i =$ $B$, $V$, $I_C$, $J$, $H$, $K_S$, $W1$, $W2$, $W3$, and $W4$
being the observed photometry, m$_{i,{\rm syn}} =$ $B_{\rm syn}$, $V_{\rm syn}$,
${I_C}_{\rm syn}$, $J_{\rm syn}$, $H_{\rm syn}$, ${K_S}_{\rm syn}$,
$W1_{\rm syn}$, $W2_{\rm syn}$, $W3_{\rm syn}$, and $W4_{\rm syn}$ are
the synthetic apparent magnitudes at the stellar surface, and $\mathcal{A}$ 
is the magnitude difference between the flux observed on Earth (obs) and
the theoretical flux at the surface of the star
(surface)\footnote{This flux is the unresolved flux integrated over
  the disk of the star and does not represent the resolved flux one
  would observe if placed on the stellar surface.  The flux we are
  referring to is the counterpart to the apparent magnitude at the
  stellar surface (e.g., $B_{\rm syn}$).}:

\begin{eqnarray*}
  \mathcal{A} & = & -2.5\log(F_{surface}/F_{obs})
\end{eqnarray*}

related to the angular semi-diameter: 

\begin{eqnarray*}
  \theta & = & \frac{R}{d} = 10^{-0.2 \mathcal{A}}
\end{eqnarray*}

We fit the observed photometry to synthetic photometry from two
different libraries of synthetic spectra: the BT-Settl
models\footnote{\url{http://phoenix.ens-lyon.fr/Grids/BT-Settl/AGSS2009/}}
of \cite{allard2012} with the \cite{asplund2009} solar composition and
the ATLAS9
models\footnote{\url{http://wwwuser.oat.ts.astro.it/castelli/grids.html}}
of \cite{castelli2004} with the \cite{grevesse1998} solar composition.
The differences in the solar composition are particularly important
for low-mass stars and brown dwarfs, due to the importance of TiO and
VO in their spectra.  The solar oxygen abundance was
revised downward by 38\% by \cite{asplund2009} compared to the
\cite{grevesse1998} oxygen abundances.  Another major difference
between the ATLAS9 models and the BT-Settl models is the treatment of
line opacities.  The ATLAS9 models include opacity distribution
functions (ODFs) to account for line blanketing, whereas the BT-Settl
models are generated by the PHOENIX code in which the individual
contribution of atoms and molecules is directly sampled over all
computed points in the spectrum \citep{hauschildt1997}.  Given that
the BT-Settl models offer continuity in our ability to model SEDs of
F-type down to M-type stars, and the recent successes the BT-Settl
models have had fitting NIR colors of low-mass stars down to
$\sim$3000~K \citep{allard2012}, we adopt the temperatures derived
from the BT-Settl models with the \cite{asplund2009} abundances, but
include the results from the ATLAS9 models to demonstrate the size of
the systematic differences resulting from the assumed solar
composition or model implementation.

\subsubsection{Testing Technique on Objects with Measured Angular
  Diameters}\label{sec:sedf}

As a reliability check for the usefulness of our method, we use the
estimated solar $BVI_C$ colors from \cite{ramirez2012} together with
the solar 2MASS $JHK$ and WISE $W1$, $W2$, $W3$ and $W4$ colors from
\cite{casagrande2012} to estimate the solar \teff, assuming
$\log(g)=4.44$ and adopting the apparent V band magnitude of
-26.74$\pm$0.02~mag \citep{mamajek2012}.  With these ten bands, the
BT-Settl models SEDF method gives \teff$_{\odot}$=5776$\pm$22~K
(remarkably within 4~K of the modern solar \teff\, of
5771.8$\pm$0.7~K; \citealt{mamajek2012}), and an angular diameter of
1949$\arcsec \pm$7$\arcsec$.  The ATLAS9 models give
\teff$_{\odot}$=5737$\pm$21~K, 35~K too low but still within
2$\sigma$, and an angular diameter of 1953$\arcsec \pm$7$\arcsec$.
Both angular diameter measurements are systematically higher than the
1918.3$\arcsec \pm$0.3$\arcsec$ angular diameter implied by the solar
radius estimate of \cite{haberreiter2008}, which strongly suggests
that our adopted $V_{\odot}$ is too high.  If we instead adopt
$V_{\odot}\equiv$-26.71$\pm$0.02~mag, we obtain angular diameters with
the SEDF method of 1922$\arcsec \pm$7$\arcsec$ and 1926$\arcsec
\pm$7$\arcsec$ with the BT-Settl and ATLAS9 models, respectively,
consistent with the modern solar angular diameter estimates.  Thus for
consistency with the solar values, also consistent with the 
\citet{engelke2010} synthetic solar V$_{\odot}$, we adopt
$V_{\odot}$=-26.71$\pm$0.02~mag\footnote{$V_{\odot}$=-26.71$\pm$0.02~mag
  implies that M$_{V,\odot}$=4.862$\pm$0.020~mag.  Based on the IAU
  scale the solar luminosity estimate of \cite{mamajek2012}
  (3.8270\,$\pm$\,0.0014 $\times$10$^{33}$ erg\,s$^{-1}$) leads to
  M$_{bol,\odot}$=4.7554$\pm$0.0004~mag,
  BC$_{V,\odot}$=-0.107$\pm$0.02~mag.  A summary of solar $V$
  magnitude estimates is available at
  \url{https://sites.google.com/site/mamajeksstarnotes/basic-astronomical-data-for-the-sun}}.

We also check our technique on nearby K- and M-type field dwarfs with
directly measured angular diameters from the recent work of
\cite{boyajian2012b}.  We use photometry from Table~7 of
\cite{boyajian2012b}, converting Johnson $I$ to the Cousins system
using the conversions in \cite{bessell1979} and converting Johnson
$JHK$ to the 2MASS system using the conversions of
\cite{carpenter2001}.  We adopt WISE $W1$, $W3$ and $W4$ photometry
with contamination and confusion flags '0' from the WISE All Sky Point
Source Catalog.  Following \cite{boyajian2012b}, we adopted
$\log(g)=4.5$ and the metallicity appropriate for each system.  We
adopt uncertainties of $\sigma_{\log(g)}$=0.2~dex and
$\sigma_{[m/H]}$=0.1~dex.  Our SEDF-derived \teff\, for these stars
are listed in Table~\ref{tbl:ad_teffs}, and plotted with the
\cite{boyajian2012b} \teff\, values in Figure~\ref{fig:teffcompare}.
The mean difference between our SEDF-derived \teff\, values and those
based on angular diameter measurements from \cite{boyajian2012b} is
13~K with a 1$\sigma$ dispersion of 108~K.  We conclude that our 
technique works well for the Sun and nearby dwarfs with angular 
diameter measurements, and gives us some confidence that this method 
will accurately predict the effective temperatures of our pre-MS stars.

\subsubsection{Analysis}

\begin{figure}
\begin{center}
\includegraphics[scale=0.45]{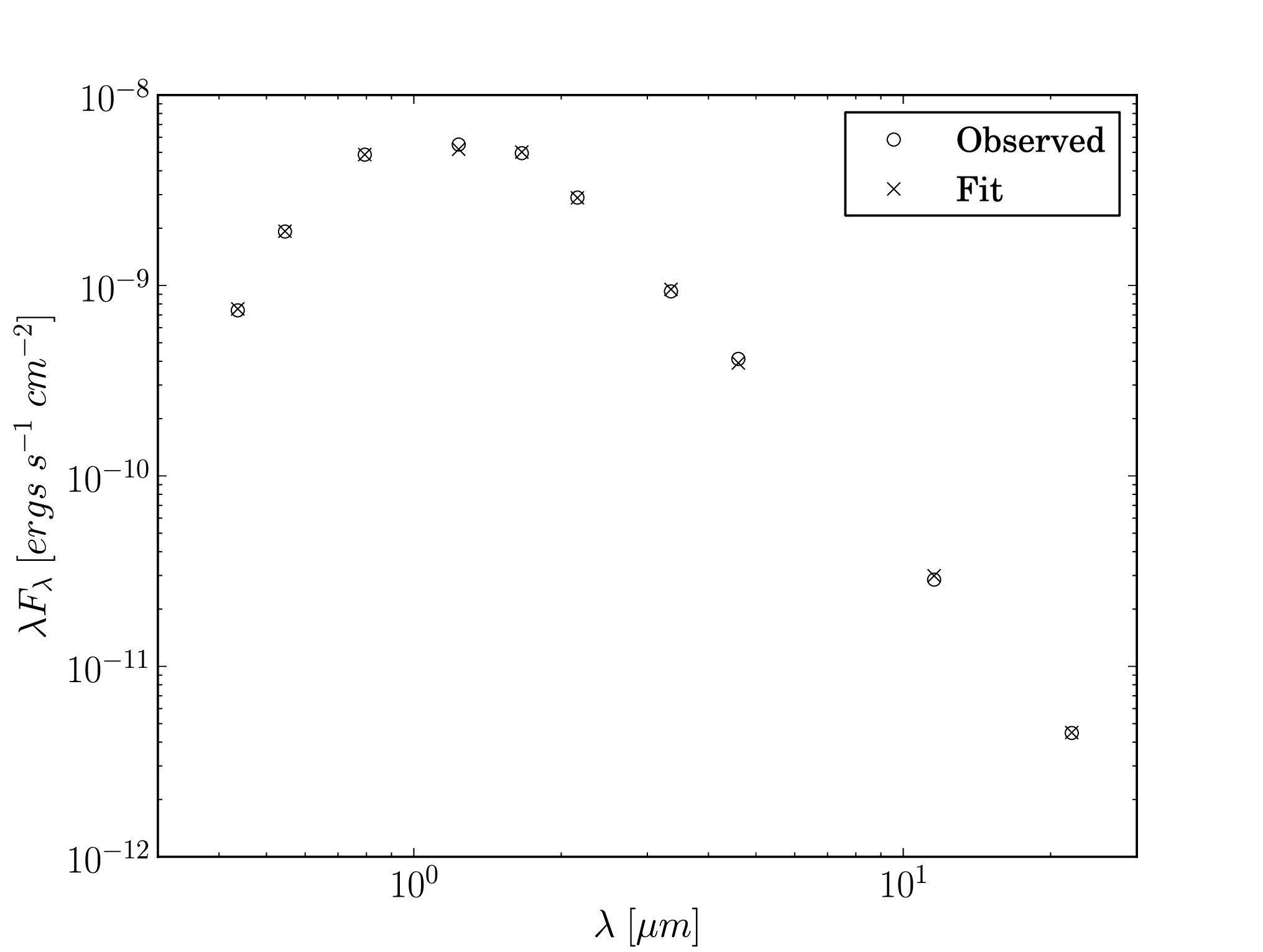}
\caption{\label{fig:sedfitcomp}
    SED for $\beta$~Pic member V1005~Ori (K8IVe).  Observed photometry 
    (circles) with the best fit BT-Settl model photometry (crosses) from a 
    \teff=3866$\pm$18~K model (interpolated).  Uncertainties are smaller 
    than the symbol markers.
}
\end{center}
\end{figure}

For many objects in our sample, one or more bands of photometry are
not available.  In those cases we simply omit the term containing the
missing band data.  We do not fit bands with poor quality photometry
(in 2MASS, anything other than quality flag `A'; for WISE bands,
anything other than contamination and confusion flag `0').  We have
again excluded photometry for objects with infrared excesses, flagged
in Table~\ref{tbl:inputdata}.  RECX~11 and RECX~15 have $K_S$-band
excesses, so we exclude them from SED fitting entirely.  We also
exclude TWA~30A due to its time variable extinction
\citep{looper2010a} and TWA~30B due to the time variable near-infrared
excess \citep{looper2010b}.  TWA~31, TWA~33 and TWA~34 have $W1$ and
$W2$-band excesses (Figures~\ref{fig:spt_color} and
\ref{fig:spt_color2}) so we exclude their WISE $W1$, $W2$, $W3$, and
$W4$ band photometry.  This leaves TWA~31 and TWA~34 with only $JHK_S$
photometry, so we exclude them entirely. TWA~29 had only 2MASS $JHK_S$
photometry, and HD~139084B and HD~164249B had 2MASS photometry
and only two bands of WISE photometry with large uncertainties ($>$
0.1 mag), which resulted in poorly constrained temperatures (e.g.,
$\sigma_{T_{\rm eff}} >$ 300~K) so we excluded them from SED fitting
as well.  Objects excluded from SED fitting are listed in
Table~\ref{tbl:sedrejects}.  The behavior of $\chi^2$ as a function of
\teff\, is consistent with Gaussian errors and $\chi^2$ has a
quadratic dependence on \teff\, near the best-fit value.  A
representative SED from our sample with the observed and best-fit
model are shown in Figure~\ref{fig:sedfitcomp}.

In general the synthetic photometry is a function of $\log(g)$, \teff,
and metallicity ([m/H]).  As discussed previously, we use solar
metallicity synthetic models.  Pre-main sequence evolutionary tracks from
\cite{baraffe1998} between 8-30~Myr predict that $\log(g)$ varies
between 4.1~dex and 4.5~dex so we simply adopt 4.3$\pm$0.2 dex.
Though it is possible to fit both \teff\, and $\log(g)$
simultaneously, this often gives spuriously large or small $\log(g)$
values, and even when the values of $\log(g)$ obtained from the fit
are within an expected range, they are not well-constrained (e.g.,
formal errors on $\log(g) \sim 1.0$ dex).  This is because most of the
synthetic colors do not depend sensitively on the adopted $\log(g)$,
and furthermore, we found that the best-fit \teff\, did not vary
significantly between $\log(g)=4.1$ and $\log(g)=4.5$.  The mean
difference in \teff\, between $\log(g)=$4.1 and 4.5 is 4~K with a
dispersion of 31~K.  Therefore, in our fitting procedure we set
\teff\, as the only free parameter.  During the fitting procedure, we
first determine $\mathcal{A}$ as the inverse-variance weighted mean
difference between the observed and synthetic photometry at the
stellar surface.  However, rather than numerically minimizing $\chi^2$
(as done in \citealt{masana2006}) we simply find the minimum value
over our grid, interpolated to \teff\, increments of 20~K from 1400~K
to 9800~K for the BT-Settl models and from 3500~K to 9750~K for the
ATLAS9 models.  We then fit a parabola in the region surrounding the
minimum.

\subsubsection{Results}

The effective temperatures from the SEDF technique are listed in
Table~\ref{tbl:teff_fit}.
We estimate our uncertainties by performing a Monte Carlo simulation.
For each object, we select trial photometry values from a distribution
with mean and standard deviation equal to the observed photometry
value and uncertainty, and use the trial photometry values to obtain
the best-fit \teff\, and angular diameter estimate.  We perform 300
trials for each object and use the standard deviation of the resulting
\teff\, and angular diameter distribution as our statistical
uncertainties.  However this does not account for systematics caused
by uncertainties in our assumed surface gravity and metallicity.  To
account for these systematics, we repeat our fitting procedure for
each object, varying $\log(g)$ from 4.1~dex to 4.5~dex and [m/H] from
+0.2~dex to -0.2~dex.  We adopt the dispersion in \teff\, and angular
diameter obtained for the systematic uncertainty, {\it typically}
$\sim$11~K in \teff\, and $\sim$1~$\mu$as in angular diameter.  The
uncertainties quoted in Table~\ref{tbl:teff_fit} are the statistical
and (internal) systematic uncertainties added in quadrature.  This
does not account for any systematic uncertainties from the underlying
Phoenix/NextGen models or the assumed solar abundances.

\begin{figure}
\begin{center}
\includegraphics[scale=0.45]{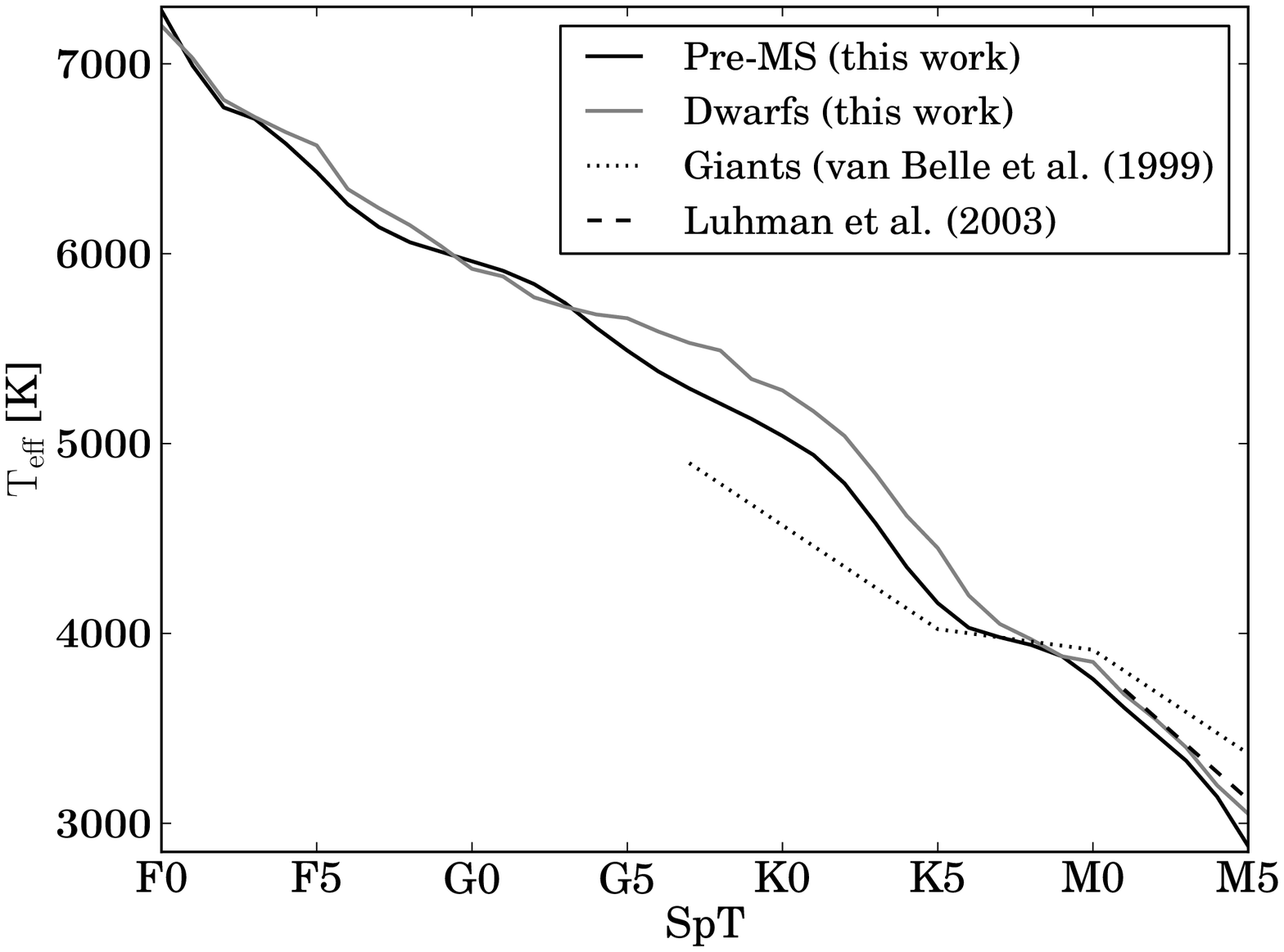}
\caption{\label{fig:spt_teff}
    Spectral type versus \teff\, for the pre-MS (solid black line) and dwarf
    (solid grey line) \teff\, scales derived in this work.  For comparison we 
    plot the giant \teff\, scale of \cite{vanbelle1999} (dotted line)
    and the \teff\, scale of \citet{luhman2003} (dashed line), appropriate 
    for $\sim$1~Myr old stars. Our pre-MS \teff\, scale is within 100~K of the 
    dwarf scale as a function of spectral type, except for spectral types G5 
    through K6, which are $\sim$250~K cooler than their main-sequence counterparts.
}
\end{center}
\end{figure}

Similar to other studies, we find that $V$--$K_S$ provides the
closest correlation to temperature with relatively little scatter.  To
take advantage of the utility of $V$--$K_S$ as a proxy for \teff, we
estimate the spectral type-temperature calibration by fitting a
polynomial to \teff\, as a function of $V$--$K_S$.  The coefficients
for this polynomial are listed in Table~\ref{tbl:teffbcpoly}.  We then
apply this polynomial to our spectral type-intrinsic color sequence.
Unfortunately only one object in our sample later than spectral type
M5.5 has $V$ band photometry, so we do not provide effective
temperature estimates for spectral types M6-M9, though we do provide
intrinsic colors for those spectral types.  Our spectral type,
intrinsic color and \teff\, sequence for young stars is listed in
Table~\ref{tbl:young_colors}.
For comparison, in Figure~\ref{fig:spt_teff} we have plotted 
the new temperature scale for 5-30 Myr pre-MS stars described in this 
work, the giant temperature scale of \cite{vanbelle1999}, a new 
``consensus'' dwarf \teff\, scale described in
Appendix~\ref{sec:dwarfcolors}, and the young star scale of
\cite{luhman2003} (appropriate for $\sim$1~Myr old stars) as a 
function of spectral type.  Our pre-MS \teff\, scale is within
$\sim$100~K of the dwarf scale as a function of spectral type,
except for spectral types G5 through K6, which are $\sim$250~K
cooler than their main-sequence counterparts.

\subsection{Bolometric Corrections}

As a byproduct of estimating the effective temperature of stars in our
sample using the method of SED fitting, we also obtain an
estimate of each object's angular diameter.  This can then be used to
estimate the apparent bolometric magnitude (m$_{\rm bol}$) and
the bolometric correction in any band ($BC_x$).  The basic
equation that relates stellar bolometric magnitude to luminosity is

\begin{eqnarray*}
  M_{\rm bol} & = & -2.5 \log\left(\frac{L}{L_{\odot}}\right) + M_{\rm bol,\odot} \\
  & = & -10 \log\left(\frac{T_{\rm eff}}{T_{\odot}}\right) - 5 \log\left(\frac{R}{R_{\odot}}\right) + M_{\rm bol,\odot} . \\
\end{eqnarray*}
We can also write this in terms of apparent magnitude $m_x$ in band
$x$ with the distance $d$ and bolometric correction $BC_x$:
\begin{eqnarray*}
  M_{\rm bol} & = & m_x - 5 \log\left(\frac{d}{10 pc}\right) + BC_x . 
\end{eqnarray*}
Equating these two, using the angular semi-diameter $\theta$ =
$\frac{R}{d}$ = 10$^{-0.2\mathcal{A}}$, and solving for $BC_x$ we find
\begin{eqnarray*}
  BC_x & = & \mathcal{A} + 5 \log\left(\frac{R_{\odot}}{10pc}\right) + M_{\rm bol,\odot} \\
       &   & - 10 \log\left(\frac{T_{\rm eff}}{T_{\rm eff,\odot}}\right) - m_x .
\end{eqnarray*}
We use consistent solar values of $T_{\rm eff,\odot} =
5772~K$, $R_{\odot} = 695660$~km, $m_{V,\odot}$ from Section~\ref{sec:sedf}, and
$M_{\rm bol,\odot} = 4.755$~mag as adopted by
\cite{mamajek2012}\footnote{See also ``Basic Astronomical Data for the
  Sun'',
  \url{https://sites.google.com/site/mamajeksstarnotes/basic-astronomical-data-for-the-sun}
  more complete discussion on solar data, including motivation for the
  values adopted here.}.  The uncertainties in $BC_x$ are
\begin{eqnarray*}
  \left(\sigma_{BC_x}\right)^2 & = & \left(\frac{10 \sigma_{T_{\rm eff}}}{T_{\rm eff} \ln10}\right)^2 + \left(\sigma_\mathcal{A}\right)^2 + \left(\sigma_{m_x}\right)^2 .
\end{eqnarray*}

\begin{figure}
\begin{center}
\includegraphics[scale=0.45]{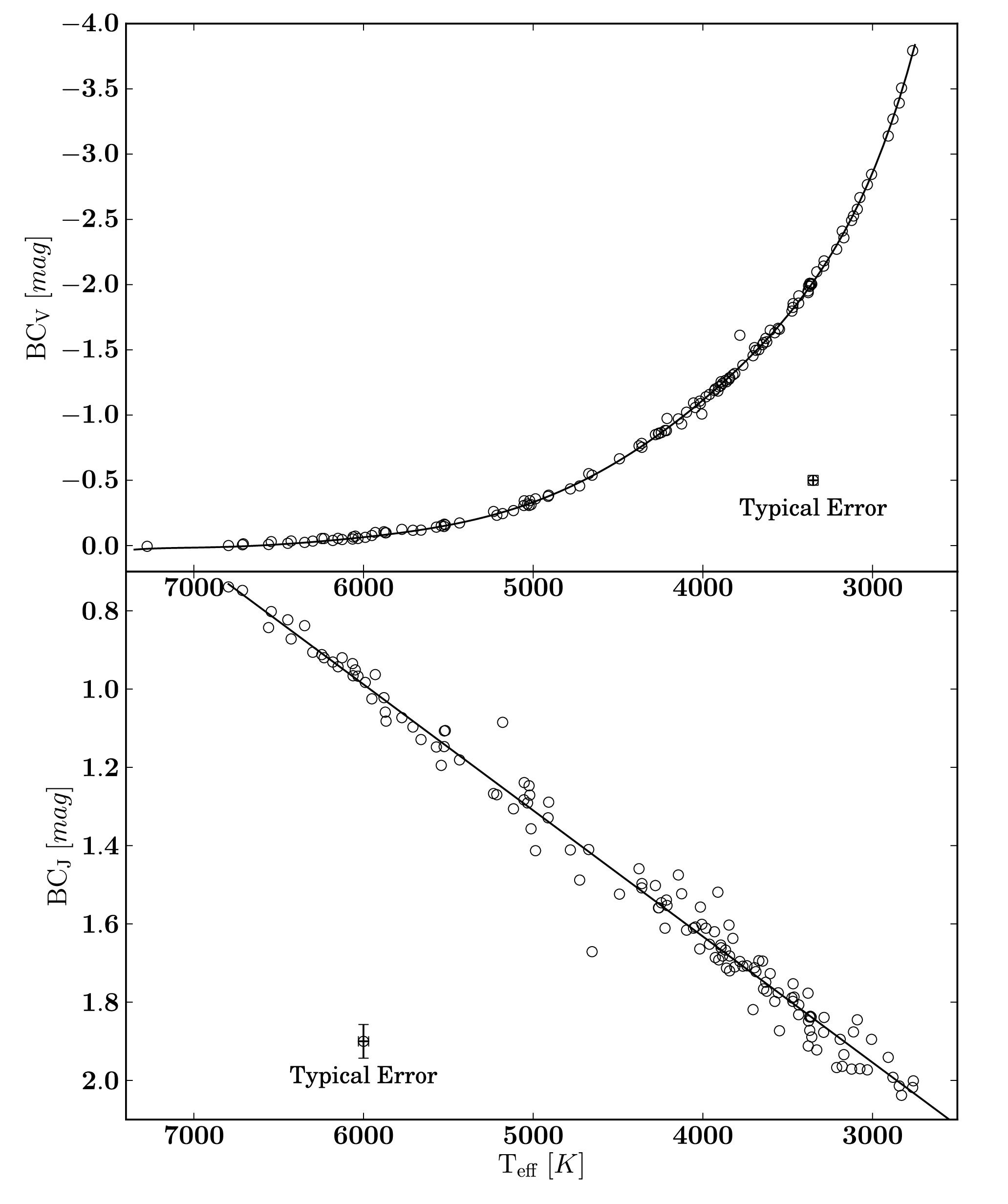}
\caption{\label{fig:teff_bc}
   Bolometric corrections for $V$ and $J$ band magnitudes as a 
   function of effective temperature.  Note that for 
   \teff $\simless$ 5000~K, $BC_V$ becomes a sensitive function
   of \teff\, and therefore it is preferable to use $M_{bol} = M_J + BC_J$
   for cooler stars.  Coefficients for polynomial fit are listed
   in Table~\ref{tbl:teffbcpoly}.
}
\end{center}
\end{figure}

Though the zero point of the bolometric correction scale is arbitrary,
the combination of bolometric correction and solar absolute
bolometric magnitude is not (see \citealt{torres2010} and Appendix~D
of \citealt{bessell1998}).  In Table~\ref{tbl:teff_fit} we give the
calculated individual bolometric corrections in both Johnson $V$ band
and 2MASS $J$ band.  We also provide $\log($L/L$_{\odot})$ for stars
with measured trigonometric parallaxes.  For F- and G-type stars
(\teff $\simgreat$ 5000~K) it is preferred to estimate bolometric
magnitudes using $M_{bol} = M_V + BC_V$ since the $V$ band correction
is not a sensitive function of \teff\, for 5000~K $<$ \teff $<$
7000~K.  However, for K- and M-type stars (\teff $\simless$ 5000~K)
$BC_V$ becomes a steep function of \teff\, and therefore it is better
to use $M_{bol} = M_J + BC_J$. Plots of $BC_V$ and $BC_J$ versus
\teff\, are shown in Figure~\ref{fig:teff_bc}.  Polynomial fits to
$BC_V$ and $BC_J$ as a function of \teff\, and $V-K_S$ are given in
Table~\ref{tbl:teffbcpoly}.

\section{Discussion}\label{sec:discussion}

Consistent with previous studies
\citep[e.g.,][]{dario2010,luhman2010}, we have found that pre-MS stars
do not have the same intrinsic colors as field dwarfs for certain
spectral types and colors.  There are two likely main reasons for the
differences in colors.  The first and most important cause is the
different surface gravities of pre-MS stars compared to main
sequence dwarfs.  The striking bifurcation in the $V$--$K_S$ versus
$J$--$H$ color-color diagram between dwarfs and giants has been
explained as an effect of CO and H$_2$O bands and H$^-$ opacity
\citep{jorgensen1996}.  The $B$--$V$ colors for pre-MS stars with
$V$--$K_S > 3.0$ are systematically bluer than main sequence stars.
At lower surface gravities, the synthetic BT-Settl $B$--$V$ colors are
predicted to be bluer at a given $V$--$K_S$ than higher-surface
gravity stars.  Our new spectral type-color relations take these
important surface-gravity effects for young stars into account.
However, this does not explain the origin of redder colors,
particularly $H$--$K_S$, for F- and G-type stars, which have surface
gravities very close to main sequence dwarfs.

The second possible explanation for color differences between young
stars and older main sequence stars suggested by \cite{gullbring1998}
and \cite{stauffer2003} is the greater abundance of stellar spots on
young stars.  Young stars show evidence of stronger magnetic activity
than older main sequence stars, which is exhibited by hotter plage and
cooler spot regions on the surface.  In particular, these plage
regions have been suggested as contributing to the systematically
bluer $B$--$V$ colors observed in the Pleiades open cluster
\citep{stauffer2003}.  \cite{gullbring1998} estimated a $\sim$50\%
spot coverage to account for the mean $V$--$J$ color anomaly in
weak-lined T Tauri stars.  However, the \cite{stauffer2003} study is
the most comprehensive attempt to date to investigate the contribution
of cool spots to stellar colors.  \cite{stauffer2003} found that
Pleiades K star red spectra (5700-8400\angstrom) had systematically
later spectral types than the blue (3300-5300\angstrom) spectra,
whereas the older Praesepe K stars did not suffer from this
effect\footnote{This effect is also seen in G and K stars from the
  younger Scorpius-Centaurus OB association, where blue spectra
  ($\sim$3800--5400\angstrom) give systematically earlier spectral
  types than the red spectra ($\sim$6200--7100\angstrom) by about one
  spectral subtype (E.E. Mamajek, private communication 2012)}.
\cite{stauffer2003} additionally modeled the $BVRIJHK$ SEDs of
several Pleiades, combining observed SEDs of an earlier field dwarf and
a later field dwarf to obtain a fit.  The best-fit models obtained in
the \cite{stauffer2003} study indicated that the K-type Pleiades were
covered in $\simgreat$50\% ``cool spots'', consistent with the
\cite{gullbring1998} results.  They use $BVRIJHK$ photometry to fit
observed Pleiad SEDs. On the basis of their spectroscopy and SED
fitting, \cite{stauffer2003} concluded that the Pleiades K stars had
more than one photospheric temperature, and that spottedness was
well-correlated with the $B$--$V$ color anomaly.  While these results
point convincingly to stellar spots as a significant contributing
factor, especially to bluer $B$--$V$ colors, we do not attempt to
quantify the relative contribution of spots or surface gravity effects
to the intrinsic colors of pre-MS stars.  Disentangling the effects of
surface gravity and spots would require time-series multi-band
photometry for most of the objects in our sample.  Quantifying the
specific contribution of the spots and plages to the stellar colors is
beyond the scope of this study.

\begin{figure}
\begin{center}
\includegraphics[scale=0.45]{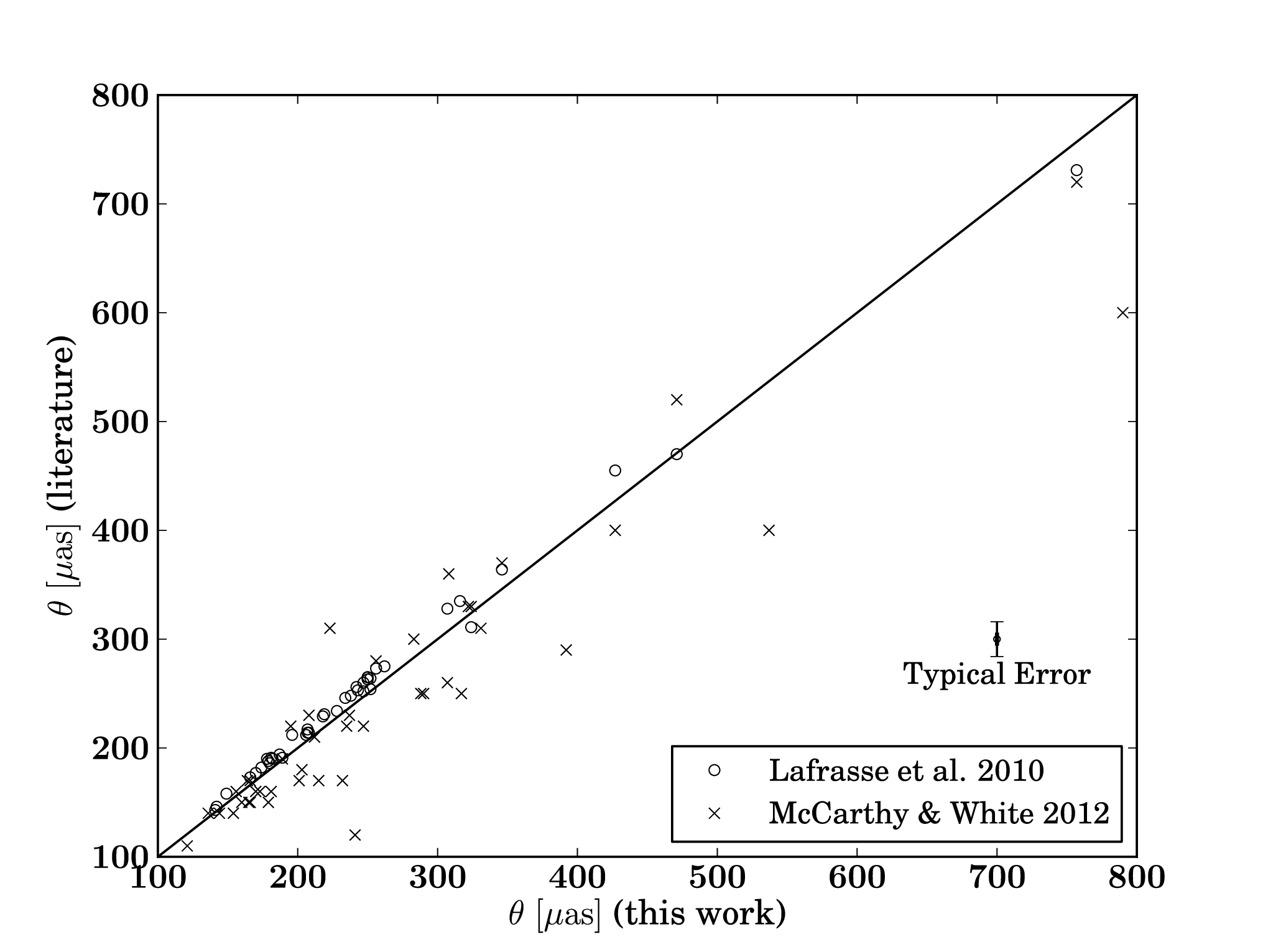}
\caption{\label{fig:diameters}
    The individual angular diameter estimates from this work compared 
    with estimates from \cite{mccarthy2012} and \cite{lafrasse2010}.
}
\end{center}
\end{figure}

\cite{mccarthy2012} published predicted angular diameters for many of
the $\beta$~Pic moving group members in our sample using estimated H-R
diagram positions and revised {\it Hipparcos} parallaxes
\citep{vanleeuwen2007}.  In addition, \cite{lafrasse2010} have
estimated the angular diameters of thousands of dwarfs and giants with
$V$ and $V$--$K$ surface brightness relations
\citep[e.g.,][]{barnes1976}.  We compare our results to the
\cite{mccarthy2012} and \cite{lafrasse2010} results in
Figure~\ref{fig:diameters}.  Our angular diameter estimates follow the
\cite{lafrasse2010} estimates very closely, though ours are
systematically smaller by 4\%.  There is a trend with \teff,
with hotter objects tend to be more discrepant than cooler objects, 
however, the origin of this discrepancy is unclear.  Our angular
diameter estimates also compare well with the results of
\cite{mccarthy2012}, with our estimates being 6\% larger on average,
but with much larger scatter, however, this difference is not statistically
significant.  The larger scatter between our angular diameter
estimates and those from \cite{mccarthy2012} are likely due to the
different methods used to infer the stellar \teffs.  For example, we
predict TYC~1208-468-1 to have a diameter of 241$\pm$1~$\mu$as, but
\cite{mccarthy2012} predict 120~$\mu$as.  This star has $BVJHK$
colors consistent with a spectral type of $\sim$K6, but it has a reported
spectral type of K3Ve \citep{jeffries1995}.  The $\sim$600~K
difference in the assumed \teff\, translates to a large difference in
the predicted angular diameter.

\begin{figure}
\begin{center}
\includegraphics[scale=0.45]{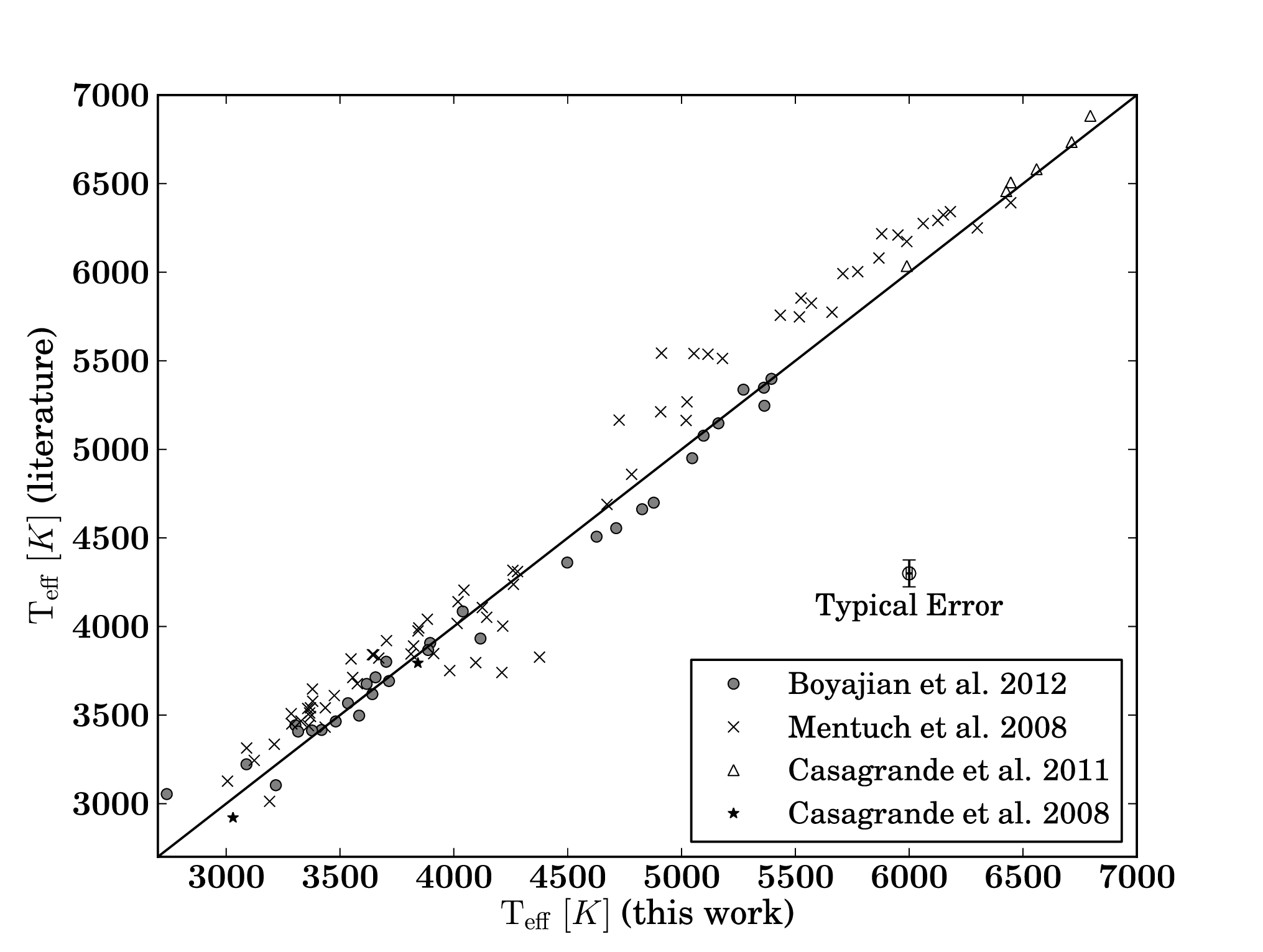}
\caption{\label{fig:teffcompare}
    The individual \teff\, values from this work compared with values
    obtained by least-squares fitting to synthetic NextGen spectra 
    from \cite{mentuch2008} (crosses) and those in the study of 
    \cite{casagrande2011} (triangles) and \cite{casagrande2008} (stars). 
    We also compare a sample of K- and M-type dwarfs which have angular 
    diameter-based \teff\, estimates from \cite{boyajian2012b} with 
    estimates using our SEDF implementation (circles).
    The values in \cite{mentuch2008} are systematically higher than 
    those estimated in this work, with the difference $\sim$230~K above 
    4500~K, reducing to $\sim$120~K below 4500~K.  Those from the 
    \cite{casagrande2011} study are typically $\sim$40~K higher than 
     the values from this work.
}
\end{center}
\end{figure}

There is considerable overlap between our sample and the sample of
\cite{mentuch2008}, who examined Li depletion in several nearby young
associations.  The \cite{mentuch2008} study estimated \teff\, for each
star in their sample by least-squares fitting synthetic spectra to 
spectral regions $\lambda\lambda$5850-5930, 7900-7980, 8150-8230, 
8400-8480, and 8485-8565 generated with NextGen model
atmospheres.  In Figure~\ref{fig:teffcompare} we compare our \teff\,
values obtained by fitting multi-band photometry to the BT-Settl
NextGen model colors with the \teff\, values obtained by
\cite{mentuch2008}.  Overall there is a systematic difference -- the
values obtained by \cite{mentuch2008} are systematically $\sim$150~K
hotter than the values we obtain, with a larger difference
($\sim$230K) above 4500~K and a smaller difference ($\sim$120~K) below
4500~K.  This discrepancy could be due to the different
synthetic models used.  The latest BT-Settl models use the revised
solar abundances from \cite{asplund2009} and include more complete
molecular opacity lists,
though these updated opacities
would mostly affect the lower-mass stars and are unlikely to account 
for the differences above $\sim$5000~K.  

In addition we have compared our estimated \teff\, values with those
of \cite{casagrande2008} and \cite{casagrande2011}, where possible 
(Figure~\ref{fig:teffcompare}).  Both studies used synthetic spectra
with an implementation of the Infrared Flux Method (IRFM) or a 
closely related method (Multiple Optical Infrared TEchnique or ``MOITE'') to
estimate the stellar effective temperature for a large number of
objects.  The IRFM compares the ratio of the observed bolometric flux 
to the observed monochromatic flux density at the Earth (``R$_{\rm obs}$'') 
to the ratio of theoretical bolometric flux to monochromatic flux density 
at the surface of the star (``R$_{\rm theo}$'') \citep{blackwell1977}.  
R$_{\rm theo}$ is a function of the \teff, and
is compared to the R$_{\rm obs}$ ratio to obtain the \teff\, of the
star.  For hotter stars the sensitivity to the model in the IR 
is very minimal and thus only these flux ratios in 
IR bands are used to determine the \teff.  For cooler stars,
\cite{casagrande2008} have adapted this method to use optical and
infrared bands (called ``MOITE'').  \cite{casagrande2008} assumed
$log(g)$=5.0~dex throughout with the `Cond' variant of NextGen models
(we have used the `BT-Settl' variant here with revised solar
abundances from \citealt{asplund2009}), whereas the
\cite{casagrande2011} study used the \cite{castelli2004} models which
used the \cite{grevesse1998} solar abundances.  Stellar \teff\, estimates 
from this work are typically $\sim$40~K lower than the values
from the \cite{casagrande2011} study (six stars in common), and within
2$\sigma$ of the values from the \citet{casagrande2008} study (stars 
TX~PsA and HIP~107345 in common).  A comparison of stellar \teff\, 
estimates from this work and the literature is shown in 
Figure~\ref{fig:teffcompare}.

For the few objects with spectral types M8 or later we obtain cooler
temperatures than expected from the temperature scale of
\cite{luhman2003} or the dwarf temperature scale.  \cite{rice2010a} 
fit PHOENIX dusty synthetic spectra to high-resolution observed 
spectra to find the best-fit \teff\, and $\log(g)$ of sample of young 
late M-type objects.  Two objects in our
sample with SEDF-determined \teff, 2MASS~J06085283-2753583 (2M0608-27;
M8.5$\gamma$; \citealt{rice2010b}) and TWA~26 (M8IVe;
\citealt{barrado2006}), are included in the \cite{rice2010a} study.
For 2M0608-27, assuming $\log(g)$=4.3~dex, we find
\teff=2118$\pm$20~K, whereas \cite{rice2010a} adopt $\log(g)$=3.98 and
\teff=2529~K, much hotter than our results and consistent with the
temperature scale of \cite{luhman2003}.  We find \teff=2176$\pm$17~K
for TWA~26 but \cite{rice2010a} find $\log(g)$=3.98 and \teff=2609~K,
again much hotter than our results and consistent with the \teff\,
scale of \cite{luhman2003}.  These four objects lack $BVI_C$
photometry and thus do not have any SED fitting constraints blueward
of their SED peak; this could be a contributing factor in their
discrepantly cool \teff\, fit.  Because of these discrepancies, we do
not include \teff\, estimates for M6 through M9 objects in our 
pre-MS temperature scale (Table~\ref{tbl:young_colors}).

Only two stars in our sample, HR~9 and 51~Eri, have direct angular
diameter measurements available from the literature.  \cite{simon2011}
measured angular diameters of 492$\pm$32~$\mu$as and 518$\pm$9~$\mu$as
for HR~9 and 51~Eri, respectively.  Our angular diameter estimates of
346$\pm$1~$\mu$as for HR~9 and 471$\pm$2~$\mu$as for 51~Eri are much
lower than the interferometric measurements.  While we have no reason
to suspect the direct measurements are unreliable {\it a priori}, the
angular diameter of 492~$\mu$as for HR~9 (F3Vn; \citealt{gray2006})
warrants some discussion.  If we adopt the estimated bolometric flux
at Earth for HR~9 from \cite{casagrande2011} of
8.6609$\times10^{-8}$~mW~m$^{-2}$, and again use the measured angular
diameter of 492~$\mu$as, we obtain a \teff\, of 5724~K, similar to a
G3V!  This is $\sim$1000~K cooler than the 6883$\pm$91~K estimated by
\cite{casagrande2011} and our estimate of 6761$\pm$28~K, both of
which are consistent with the F3Vn spectral type.  The
\cite{simon2011} results indicate a larger angular diameter at $H$
band than $K$ band, which points to unusual calibration errors
(M. Simon, private communication 2012).  We suspect that our predicted
angular diameters are closer to the actual diameters and until updated 
measurements are published, we recommend our predicted angular diameter.

\section{Conclusions}

We can summarize our conclusions as follows:

\begin{enumerate}

\item 5-30~Myr old pre-main sequence stars follow slightly different
  spectral type-intrinsic color sequences than that of main sequence
  stars.  Pre-MS colors follow the dwarf sequence for some colors and
  spectral types, but for other optical/infrared colors and spectral 
  types, deviations can exceed 0.3~mag.  In Table~\ref{tbl:young_colors} 
  we provide an empirical tabulation of the intrinsic colors of young 
  stars for spectral types
  F0 through M9, including $B$-$V$, $V$--$I_C$, $V$--$K_S$, $J$--$H$,
  $H$--$K_S$, $K_S$--$W1$, $K_S$--$W2$, $K_S$--$W3$, and $K_S$--$W4$.

\item Consistent with previous findings \citep{luhman1999b,dario2010}, 
  we find that color differences between K- and M-type pre-MS stars and 
  dwarfs appear to be predominantly due to the young stars' lower surface 
  gravities. This is demonstrated by theoretical models predicting redder 
  $J$--$H$ colors and bluer $B$--$V$ colors for lower surface gravity 
  objects, consistent with observations.  However, we cannot exclude hotter 
  plage and cooler spot regions on the stellar surface as contributing factors.

\item A pre-MS \teff\, scale derived from fitting SEDs to synthetic
  spectral models is within $\sim$100~K of main sequence stars as a
  function of $V$--$K_S$.  As a function of spectral type, the
  effective temperatures of F0 through G4 and K7 through M5 pre-MS stars are within
  $\sim$100~K of their main sequence counterparts, whereas G5 through K6
  pre-MS stars are $\sim$250~K cooler at a given spectral type.  We
  provide new spectral type-\teff\, relations and color-\teff\,
  relations appropriate for 5-30~Myr old pre-MS stars.  We also
  provide bolometric corrections appropriate for PMS stars as
  polynomial functions of \teff\, and $V$--$K_S$ in
  Table~\ref{tbl:teffbcpoly} and as part of our spectral
  type-intrinsic color sequence in Table~\ref{tbl:young_colors}.

\end{enumerate}

We thank Fred Walter for the use of his IDL reduction pipeline and the
Stony Brook Spectral Standards Library. We also thank the referee,
Kevin Luhman, for his very thorough and prompt review which greatly
improved the paper. Spectra taken for this study were observed on the
1.5-m telescope on Cerro Tololo via the Small and Moderate Aperture
Research Telescope System (SMARTS) Consortium.  
We thank Duy Nyugen for helpful discussions regarding $\chi^2$
fitting.
This work was supported by funds from NSF grant AST-1008908.
This publication makes use of data products from the Two Micron All
Sky Survey, which is a joint project of the University of
Massachusetts and the Infrared Processing and Analysis
Center/California Institute of Technology, funded by the National
Aeronautics and Space Administration and the National Science
Foundation.
This publication makes use of data products from the Wide-field
Infrared Survey Explorer, which is a joint project of the University
of California, Los Angeles, and the Jet Propulsion
Laboratory/California Institute of Technology, funded by the National
Aeronautics and Space Administration.
This research was made possible through the use of the AAVSO
Photometric All-Sky Survey (APASS), funded by the Robert Martin Ayers
Sciences Fund.
This research used the Digitized Sky Survey, NASA ADS, and the SIMBAD
and Vizier databases.

\bibliography{ms}

\LongTables

\appendix
\section{Membership of TWA 9 to the TW Hya
  Association}\label{sec:twa9}

The membership of TWA~9 to the TW Hya Association merits some
discussion.  \cite{weinberger2012} showed that the space motion of
TWA~9A is more than 3$\sigma$ from the mean of the association, and
concluded that it was either not a member or the {\it Hipparcos}
distance is underestimated.  However, when considering the TWA
centroid space motion \citep{weinberger2012}, the Tycho-2 proper
motion ($\mu_{\alpha}*$=-55.4$\pm$2.3~mas~yr$^{-1}$,
$\mu_{\delta}$=-17.7$\pm$2.3~mas~yr$^{-1}$; \citealt{hog2000}) of
TWA~9A seems consistent with membership in TWA.  Assuming it is a
member and adopting the TWA mean group space motions from
\cite{weinberger2012} of (U, V, W) = (-10.9$\pm$0.2, -18.2$\pm$0.2,
-5.3$\pm$0.2)~km~s$^{-1}$, we estimate a kinematic distance of 
70.0$\pm$3.8~pc, based on the method discussed in \citet{mamajek2005}.  
If we adopt this kinematic distance with the
\citealt{bailey2012} mean radial velocity for component A and B of
11.964$\pm$0.024~km~s$^{-1}$, the 3-D space motion of TWA~9A is then
(U, V, W) = (-10.2$\pm$1.2, -19.7$\pm$0.8, -4.8$\pm$0.6)~km~s$^{-1}$.
This is consistent with the mean TWA space motions in the
\cite{weinberger2012} study.  Furthermore, the kinematic distance
would decrease the absolute magnitude $M_H$ by $\approx$0.83~mag over
the {\it Hipparcos} distance (using d=$\pi^{-1}$, where $\pi$ is the
trigonometric {\it Hipparcos} parallax), and thus the isochronal age
of TWA~9A would be $\sim$16~Myr, much closer to the isochronal ages
obtained by \cite{weinberger2012} for other TWA members.  TWA~9A
exhibits very high Li (EW(Li~6708\angstrom)=470~m\angstrom;
\citealt{torres2006}), lies in the direction of other TWA members, has
proper motion consistent with membership in TWA, and, adopting the
kinematic distance of 70.0$\pm$3.8~pc, has a space motion and
isochronal age consistent with membership in TWA.  Thus, we retain
TWA~9A and TWA~9B as members of TWA and suggest that the {\it
  Hipparcos} parallax is most likely $\sim$3$\sigma$ in error.

\section{Spectral Transition from K7 to M0}\label{sec:k7_m0}

Some spectral surveys implicitly or explicitly do not recognize or use
spectral types K8 and K9.  While spectral types K8 and K9 are not
considered full subtypes of the spectral classification system
\citep{keenan1984}, it should be pointed out that neither are G1, G3,
G4, G6, G7 or G9, yet these classifications are consistently
recognized and used \citep[e.g.,][]{gray2003}.  \cite{keenan1984}
noted that subdivisions such as G3 simply means the star is closer
to G2 than G5, and that they should be used when it is possible to
classify the stars accurately enough to justify their use.
\cite{keenan1984} considered K5 and M0 one subtype apart even though
the difference in their $B$--$V$ color is 0.3~mag, larger than the
difference between M0 and M4 (see Table~\ref{tbl:ic_teff}).  From the
standpoint of spectral classification, there is nothing different
about the K7 to M0 transition that merits such a gap in spectral
types.  Therefore we find no compelling reason to omit spectral types
K8 and K9 from use and we include them here in our analysis.

With low-resolution red optical spectra we can distinguish between
subtypes K7, K8, K9 and M0.  Unfortunately, K8V and K9V spectral
standards do not appear in the literature (e.g., \citealt{gray2009}).
For these subtypes, we adopted stars as standards which were assigned
this classification by an expert classifier.  For K8V we adopted
HIP~111288 ($V$--$K_S$=3.52$\pm$0.02~mag;
\citealt{perryman1997,skrutskie2006}), classified as K7V by
\cite{stephenson1986} but later classified as K8Vk by \cite{gray2006}.
For K9V we adopted HIP~3261 ($V$--$K_S$=3.70$\pm$0.02~mag;
\citealt{perryman1997,skrutskie2006}), classified as K7.0 by
\cite{hawley1996} but later classified as K9V by \cite{gray2006}.
These were chosen because they were classified as intermediate between
K7 and M0 but also because they have $V$--$K_S$ colors intermediate
between K7 and M0.  We visually compared the spectra of both adopted
standards and verified that they were morphologically intermediate
between the K7V and M0V standards in Table~\ref{tbl:spstands}.

\begin{figure}
\begin{center}
\includegraphics[scale=0.45]{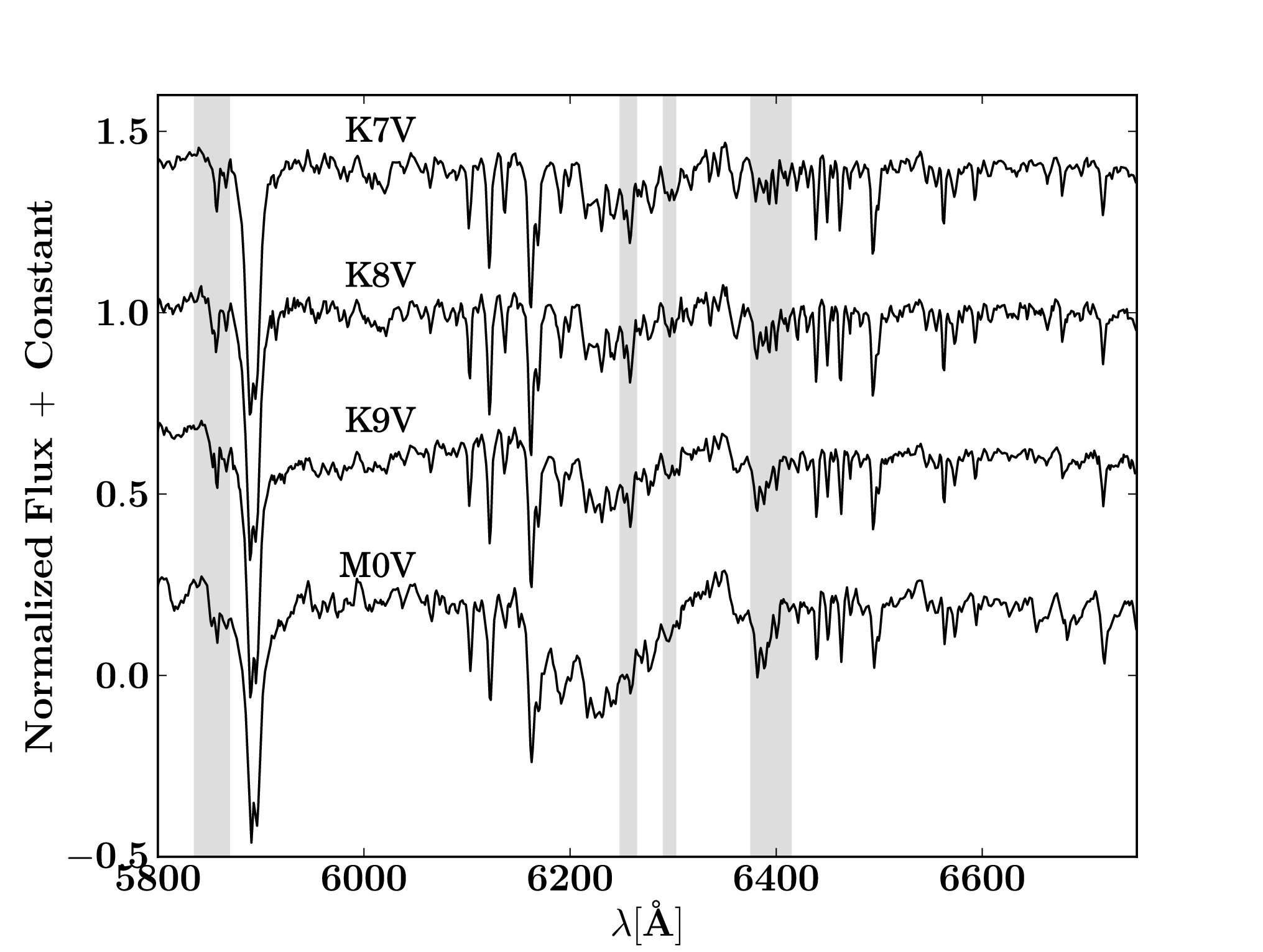}
\caption{\label{fig:k7_m0}
    The spectral transition from K7 to M0, with regions most useful for
    discriminating among the different spectral types highlighted in 
    grey.  GJ~673 (K7V), HIP~111288 (K8Vk), HIP~3261 (K9V) and GJ~701 
    (M0.0V) are shown.
}
\end{center}
\end{figure}

Figure~\ref{fig:k7_m0} shows the red spectral sequence from K7 to
M0.  The spectra show a distinct progression in the CaH band at
$\lambda\lambda$ 6382-6389 \citep{kirkpatrick1991, torres-dodgen1993,
  allen1995} which gradually becomes stronger than the Fe~I lines at
6400\angstrom\, and 6393\angstrom.  Another very useful discriminant
is the relative strength of the V, Ti, and Fe blend at 6297\angstrom\,
to the Fe~I blend at 6302\angstrom.  As the TiO band
$\sim$6080-6390\angstrom\, increases in strength going from the late
K-types to M0, the relative strength of these two lines gradually
changes, with the blend at 6297\angstrom\, becoming stronger at K8.

Lastly, we mention that skipping from K7 to M0 may hide important
discrepancies between models and observations.  \cite{casagrande2008}
has noted that, among main sequence dwarf stars, there appears to be a
plateau in luminosity in the transition region between K and M where
the stars appear to be decreasing in temperature with very little
decrease in luminosity.  Theoretical models do not appear to reproduce
the plateau.  Using subtypes K8 and K9, when possible, presents an
opportunity for observational work to make contact with theoretical
models in this region.

\section{Dwarf Colors and Temperatures}\label{sec:dwarfcolors}

In order to accurately compare the empirical intrinsic colors and
\teff\, scale of pre-MS stars to dwarfs and quantify their
differences, the empirical intrinsic colors of dwarf stars must be
accurately tabulated.  Here we describe the construction of a modern
dwarf color, \teff, and bolometric correction sequence, which has been
an ongoing process carried out over several years. While other
compilations are available \citep[e.g.][]{schmidt-kaler1982,
  kenyon1995, worthey2011}, it was our goal to incorporate a detailed
review of the color/temperature placement of modern spectral standard
stars and assess their pedigree as standards.  A preliminary version
of this sequence (A0V-G9V) was previously published in
\citet{pecaut2012}. The primary motivations for constructing this
sequence were that (1) color sequences over the O-M range of spectral
types had not been constructed explicitly including 2MASS and WISE
bands, (2) the methodology of the construction of previous sequences
was not always made clear, (3) systematic differences exist between
some of the widely cited past sequences, (4) there have been sizable
shifts in \teffs\, reported for some stars (especially among the
hottest and coolest dwarf stars) over the past few decades, and (5)
there have been subtle changes to the dwarf spectral sequence over the
decades, especially among the M dwarfs. In light of the subtle shifts
of the MK system over the past decades, improvements in the modeling
of stellar atmospheres, and given the large volume of optical-IR
photometry and derived stellar parameters in the literature now, a
reevaluation of the temperature and colors scales is overdue.

We present our modern intrinsic color-\teff-spectral type tabulation
for dwarfs in Table~\ref{tbl:ic_teff}.  This color tabulation was
independently derived, and is {\it not} dependent on previous
compendia of dwarf photometric properties. There were several stages
that went into assembling Table \ref{tbl:ic_teff}.  When discussing
samples of ``nearby stars'', we assumed that stars with trigonometric
parallax distances within 75 pc had negligible reddening
\citep[e.g.][]{reis2011}, and so could be used to estimate intrinsic
colors.  While we often quote the intrinsic stellar colors to 0.001
mag precision (to ensure construction of smooth sequences on
color-color plots), the uncertainty in the mean colors is typically at
the $\sim$0.01 mag level, but can be upwards of a few hundredths of a
magnitude for the M-type dwarfs (where the uncertainties reflect
differences among the colors of the standard stars themselves, and in
the mean colors for stars of a given subtype). The first step was to
estimate the mean intrinsic colors for each dwarf spectral subtype
among one or more colors sensitive to spectral type, using both
standard stars and samples of field stars with subtypes measured by
expert classifiers. To anchor spectral type to these colors sensitive
to spectral type, we used dereddened U-B colors for OB dwarf stars,
B-V colors for AFGK dwarfs, and V-K$_s$ colors for M dwarfs.

\subsubsection{Spectral Standard Stars}

 Spectral standard stars for stars of F-type and earlier were mostly
 drawn from \citet{johnson1953, morgan1965, garrison1967, lesh1968,
   abt1968, hiltner1969, cowley1969, garrison1972, cowley1972,
   morgan1973, cowley1974, houk1975, garrison1977, morgan1978,
   garrison1979, gray1987, walborn1990, garrison1994, garrison1994MK,
   gray2009}. For M-type stars, the primary sources of standard stars
 were \citet{kirkpatrick1991, kirkpatrick1997, henry2002}.  Some
 M-type standards from Keenan's papers \citep[e.g.][]{keenan1976,
   keenan1980, keenan1983, keenan1988, keenan1989} that have
 conflicting types compared to the newer classifications by
 Kirkpatrick, Henry, and collaborators, have been deprecated (e.g. GJ
 15A, 172, 250B, 526) and were not considered in assessing median
 colors and \teff. Given the immense volume of recent M-star
 classifications that have been done on the Kirkpatrick \& Henry grid
 \citep[e.g.][]{reid1995, hawley1996, henry2002}, these should be
 preferred to the Keenan types where there is
 disagreement. Classifications of AFGK field dwarfs by
 \citet{gray2003, gray2006} were generally preferred over those of the
 Michigan Atlas \citep[][]{houk1975}, as it appears that the Gray et
 al. classifications more closely follow the Morgan-Keenan
 standards. Differences between Gray et al. and Houk et
 al. classifications are especially pronounced amongst the early
 G-type stars. Part of this may stem from disagreement between Morgan
 and Keenan on the F/G boundary (e.g. see the example of $\eta$ Cas A
 previously mentioned). More problematically, \citet{houk1975}
 considered $\beta$ Com to be their main G2V standard, but it was
 considered to be a G0V standard by \citet{johnson1953, morgan1971,
   morgan1973, keenan1976}. This appears to explain why the median B-V
 color for nearby G2V stars in the Hipparcos catalog (dominated by
 Michigan Atlas classifications) is B-V $\simeq$ 0.617, whereas for
 stars classified G2V by \citet{gray2001a, gray2003, gray2006} on
 Keenan's standard star grid, B-V $\simeq$ 0.647 \citep[remarkably
   similar to the recent precise estimate of the solar B-V color of
   0.653\,$\pm$\,0.003 by ][]{ramirez2012}. As Keenan's G/K-type
 standard stars are in common usage, we weight the median colors for
 field dwarfs classified using the Keenan standards (e.g. Gray's
 papers) over those from the Michigan Atlases
 \citep[e.g.][]{houk1975}.

\subsubsection{Assessing the Pedigree of Spectral Standard Stars}

An extensive literature search was conducted to assemble notes on the
published classifications and colors for {\it all} known O- through
M-type dwarf spectral type standard stars (The notes have been
compiled at \url{http://www.pas.rochester.edu/\~emamajek/spt/} and
will be periodically updated as needed).  All the dwarf spectral
standards were assessed for continuity in their spectral
classifications over the decades, and standards were graded as
``anchor standards'' \citep{garrison1994MK}, ``primary standards'',
``secondary standards'', ``tertiary standards'', ``variant
standards'', or ``deprecated standards''. Our terminology is a
variation on the hierarchy scheme of \citet{garrison1994MK}, and the
goal of assessing the pedigree of the various spectral standards was
to help in the estimation of the best stellar parameters reflective of
a given spectral subtype.  While the grading of the individual
standards is not provided here, the reader is referred to the website
mentioned. 
``Anchor standards'' are those rare standard stars listed by
\citet{garrison1994MK} whose spectral types have remained unchanged
since \citet{morgan1943}, and which essentially define the MK system.
``Primary standards'' typically showed very strong continuity in
adopted spectral types among expert classifiers, often going back to
\citet{johnson1953}. 
``Secondary standards'' usually appeared several times in the
literature as spectral standards, but sometimes expert classifiers
assigned slightly different spectral types to the star (usually at the
$\pm$0.5-2 subtypes level).
``Tertiary standards'' were rarely graded as such, but this was
usually the category assigned when the standard was only considered as
such by one study, and with no or few dissensions or corroborating
classifications, e.g. the B8V standard HR 9050 (considered only a
standard by \citealt{garrison1994}).
``Variant standards'' are standards with spectral peculiarities
(usually demonstrating very non-solar composition), e.g. the G5Vb Fe-2
star 85~Peg \citep{keenan1988}, and these were ignored when
considering adopted subtype colors.  As \citet{garrison1994MK}
discussed, the ``anchor standards'' are those that have remained
unchanged since \citet{morgan1943}, and essentially define the
backbone of the MK system. Occasionally, a star whose classification
has varied over the years is considered a primary standard only
because no better standard is available \citep[e.g. 16 Cyg B, a
  ``primary'' G3V standard, but whose classifications have varied from
  G2V to G5V over the years;][]{keenan1976, keenan1989, gray2001a}.
``Deprecated standards'' were considered those standard stars whose
spectral types determined by expert classifiers had changed
appreciably over the years (even by the same classifier!), while higher
pedigree standards for that subtype were available.  An example of a
deprecated standard is $\eta$~Cas~A, considered an F9V standard by
\citet{keenan1988, keenan1989} and \citet{gray2001a}, but considered a
G0V standard by \citet{morgan1943, johnson1953, morgan1973,
  keenan1976, morgan1978, keenan1983}, and \cite{keenan1985}.  Another
example is $\sigma$ Boo (HR 5447), which was considered a F2V standard
by \citet{morgan1943} and \citet{johnson1953}, but two later studies
found the star to appear spectrally metal poor \citep[F3V
  vw;][]{barry1970} and \citep[F4V kF2 mF1;][]{gray2001a}. Use of such
standards should probably be avoided in the future, if possible.\\

While estimating the parameters for a given dwarf spectral subtype,
more weight was assigned to the individual parameters (e.g. colors,
\teffs) of the anchor and primary standards compared to the secondary
and tertiary standards, and the properties of the variant and
deprecated standards were largely ignored. While estimating the
typical properties of non-standard stars of a given spectral subtype,
we employed median values throughout, in order to avoid the effects of
interloper data \citep{gott2001}. The properties of both standard and
non-standard stars were incorporated into estimation of typical colors
and \teffs, and their properties usually agreed well with very few
exceptions (e.g. B7V, where the lone good standard star HD 21071
appears to be significantly bluer and hotter than the majority of
field stars classified B7V).

\subsubsection{Color Sequences}

The intrinsic ($B$--$V$)$_o$ and ($U$--$B$)$_o$ colors can be derived
for OB dwarfs via the $Q$-method
\citep[e.g.,][]{johnson1953,johnson1958,hiltner1956}, where the
reddening-free index $Q$ is calculated using the observed colors as
$Q$=($U$--$B$)-0.72($B$--$V$).  Functions of ($B$--$V$)$_o$ and
($U$--$B$)$_o$ as linear functions of $Q$, especially those that are
forced through the origin (($B$--$V$)$_o$, ($U$--$B$)$_o$), produce
poor fits to the colors of real unreddened OB stars. We calibrated new
$Q$ versus intrinsic color relations using $UBV$ photometry from
\cite{mermilliod1994} of nearby negligibly reddened B-type dwarfs
within 75~pc ({\it Hipparcos} catalog; \citealt{esa1997}), and lightly
reddened hotter O and early-B luminosity class V and IV stars in
nearby associations. The more distant OB stars were dereddened using
published H~I column densities \citep[e.g.,][]{fruscione1994} and the
strong correlation between N(H~I) and $E$($B$--$V$);
\cite{diplas1994}. The improved Q-method fits are:

\begin{eqnarray*}
(B-V)_o  & = & -4.776008156728\times10^{-3} + \\
         &   & 0.5522012574154\,Q + \\
         &   & 1.151583004497\,Q^2 + \\
         &   & 1.829921229667\,Q^3 + \\
         &   & 0.8933113140506\,Q^4 \\
\end{eqnarray*}

for $-0.32 < (B-V)_o < 0.02$, and

\begin{eqnarray*}
(U-B)_o  & = & 6.230566666312\times10^{-3} + \\
         &   & 1.533217755592\,Q + \\
         &   & 1.385407188924\,Q^2 + \\
         &   & 2.167355580182\,Q^3 + \\
         &   & 1.075207514655\,Q^4
\end{eqnarray*}

for -1.13 $<$ ($U$--$B$)$_o$ $<$ 0.02. We find that the intrinsic
($B$--$V$)$_o$ colors of O9/B0 dwarfs are -0.32 to -0.31 (among the
calibrator stars e.g. 10 Lac, $\sigma$ Sco, $\tau$ Sco, and $\upsilon$
Ori), in agreement with Johnson's classic work
(e.g. \citealt{johnson1953,johnson1966}, but at odds with the recent
work of \cite{martins2006} who claim that ($B$--$V$)$_o$ colors of
Galactic O stars go no bluer than -0.28.

Deriving the main sequence color sequence was fairly straightforward.
Photometry for nearby stars came from the following sources: $UBV$
\citet{mermilliod1991}, $BVI_C$ \citep{esa1997}, $JHK_S$
\citep{skrutskie2006}, $W1$, $W2$, $W3$, and $W4$ \citep{cutri2012}.
While we did derive median color estimates for each type (some are
listed in the individual spectral type files), we decided to fit
polynomials to the color-color data for nearby field dwarfs. For some
color-color sequences polynomials give inadequate fits. For these
instances we found it more reliable to simply construct a well-sampled
color-color table based on median colors within a given color bin, and
interpolate (e.g. $V$--$K_S$ vs. $B$--$V$, $V$--$I_C$, $J$--$H$,
$H$--$K_S$, $B$--$V$ vs $V$--$I_C$ and $U$--$B$).  We fit polynomial
relations to $V$--$K_S$ versus $K_S$--$W1$, $K_S$--$W2$, $K_S$--$W3$,
and $K_S$--$W4$ for stars within 75~pc from the {\it Hipparcos}
catalog and the catalog of bright M dwarfs from \citet{lepine2011}.
We adopted $V$ magnitudes from the APASS Data Release 6 catalog
\citep{henden2012} for objects not present in the {\it Hipparcos}
catalog, and only fit objects with high quality photometry in the
relevant band (for 2MASS bands, quality flag `A'; for WISE bands,
contamination and confusion flag `0').  We restricted the data to WISE
magnitudes $W1 > 5.0$, $W2 > 6.0$, $W3 > 5.0$ and $W4 > 0.0$ to avoid
biases due to saturation.  The data is not well-populated for
$V$--$K_S < 0.5$~mag or $V$--$K_S > 6.0$~mag, so
Table~\ref{tbl:ic_teff} only contains WISE colors for spectral types
F0 through M5.

\subsubsection{Effective Temperatures}

Subtype \teffs\, were estimated by considering published \teffs\, for
individual stars of a given subtype, though greater weighting was
given to \teff\, values for spectral standards which were vetted for
consistent classifications in the literature.  Our search for
published \teffs\, was extensive, though not exhaustive, and given
time constraints we are admittedly limited by what values were
published in electronic tables that could be easily queried with
e.g. Vizier\footnote{\url{http://vizier.u-strasbg.fr/viz-bin/VizieR}}. Many
\teffs\, came from large catalogs by e.g.  \citet{philip1980,
  sokolov1995, cayrel1997, blackwell1998, gray2001b, gray2003,
  taylor2005, valenti2005, paunzen2006, gerbaldi2007, fitzpatrick2007,
  prugniel2007, zorec2009, soubiran2010, casagrande2011}, and the
authors calculated photometric \teffs\, for OB dwarf standards using
photometry from \citet{hauck1998}, dereddening relations from
\citet{castelli1991}, and color-temperature relations from
\citet{balonashobbrook1984, napiwotzki1993, balona1994}.  \teffs\,
were also estimated for OB dwarf standards using $U$--$B$ vs. \teff\,
data in \citet{bessell1998}.

Here is an example of our evaluation of the median \teff\, for A0V
stars. We find very consistent effective temperatures among A0V
standards within a few hundred K of each other. The A0V standard Vega
has had a very precise apparent \teff\, measured by
\citet{monnier2012} of 9660\,K (in good agreement with many previous
estimates), and we find the literature median \teff\, for the other
widely used MK standards $\gamma$ UMa and HR 3314 to be 9361\,K and
9760\,K.  While there are other A0V standards, two of these ($\gamma$
UMa, Vega) are considered ``anchor'' standards by \citet{garrison1994MK}
(i.e. their classifications have remained the same over seven decades of use),
and HR 3314 has retained its A0V standard status throughout
\citep{morgan1953, johnson1953, garrison1979, gray1987, houk1999,
  gray2003}. An exhaustive search for \teffs\, for A0V stars in the
literature (265 estimates) yields a median \teff\, of 9707\,K. Based on
these values, we adopt a median \teff\, of 9700\,K for A0V stars. We
find it unlikely that the median A0V \teff\, could be as high as
10000\,K \citep{bessell1979, crowther1997}, nor as low as 9394\,K
\citep{boyajian2012a}, 9520\,K \citep{schmidt-kaler1982}, or 9530\,K
\citep{theodossiou1991}. We note in particular that the recently
published \teff\, scale by \citet{boyajian2012a} appears to be most
deviant among the A0V \teff\, values, and while that study relies on
new interferometric observations, their survey contained only a single
non-standard A0V star (HD 177724). Similarly sized deviations at the
hundreds of K level were seen between our \teff\, scale and the
\citet{boyajian2012a} \teff\, scale. So while there are other modern
color/\teff\, scales in the literature, we believe that ours is based
on a very broad (but vetted) amount of photometric/\teff\, literature
and classifications.

\subsubsection{Bolometric Corrections}

The bolometric corrections (BCs) listed in Table~\ref{tbl:ic_teff} are
derived for each spectral type by adopting the median BC among several
scales as a function of the adopted \teff, including
\citet{balona1994, bertone2004, flower1996, bessell1998, masana2006,
  schmidt-kaler1982, code1976, casagrande2006, casagrande2008,
  casagrande2010, lanz2007, vacca1996, lanz2003}\footnote{Extensive
  notes and discussion can be found for each spectral type at
  \url{http://www.pas.rochester.edu/~emamajek/spt/}} where they
applicable. For the M dwarfs, the BC$_V$ scale was estimated via
V-K$_s$ colors and the BC$_K$ results from \citet{Leggett01},
\citet{dahn2002}, and \citet{golimowski2004}, as well as the authors' SEDF
fits compiled in Table~\ref{tbl:ad_teffs}. 

\clearpage
\begin{landscape}
\begin{deluxetable}{lllllrrrrrrrrrrrr}
\tabletypesize{\scriptsize}
\tabletypesize{\Tiny}
\setlength{\tabcolsep}{0.02in}
\tablewidth{0pt}
\tablecaption{Spectral Types and Optical/Near-IR Photometry for Young, Nearby, Moving Group Members.\label{tbl:inputdata}}
\tablehead{
  \colhead{Star} &
  \colhead{Grp} &
  \colhead{2MASS} &
  \colhead{SpT} & 
  \colhead{Ref.} & 
  \colhead{$V$} & 
  \colhead{$B-V$} & 
  \colhead{$V-I_C$} & 
  \colhead{Ref.} &
  \colhead{$J^a$} & 
  \colhead{$H^a$} & 
  \colhead{$K_S^a$} & 
  \colhead{$W1^b$} & 
  \colhead{$W2^b$} & 
  \colhead{$W3^b$} & 
  \colhead{$W4^b$} \\
  \colhead{} & 
  \colhead{} & 
  \colhead{} & 
  \colhead{} & 
  \colhead{} & 
  \colhead{(mag)} & 
  \colhead{(mag)} & 
  \colhead{(mag)} & 
  \colhead{} &
  \colhead{(mag)} & 
  \colhead{(mag)} & 
  \colhead{(mag)} & 
  \colhead{(mag)} & 
  \colhead{(mag)} & 
  \colhead{(mag)} & 
  \colhead{(mag)} 
}
\startdata
HIP~490 & TH & 00055255-4145109 & G0V & 1 & 7.510$\pm$0.010 & 0.595$\pm$0.008 &  & 2,2 & 6.464$\pm$0.020 & 6.189$\pm$0.023 & 6.117$\pm$0.020 & 6.043$\pm$0.053 & 6.053$\pm$0.023 & 6.105$\pm$0.015 & 5.975$\pm$0.041 \\
HR~9 & BP & 00065008-2306271 & F3V & 3 & 6.190$\pm$0.010 & 0.386$\pm$0.007 &  & 2,2 & 5.451$\pm$0.024 & 5.331$\pm$0.047 & 5.240$\pm$0.024 & 5.245$\pm$0.072 &  & 5.234$\pm$0.014 & 4.514$\pm$0.023$^\dagger$ \\
HIP~1113 & TH & 00135300-7441178 & G6V & 4 & 8.760$\pm$0.010 & 0.752$\pm$0.021 & 0.820$\pm$0.010 & 2,5,6 & 7.406$\pm$0.021 & 7.087$\pm$0.029 & 6.962$\pm$0.023 & 6.888$\pm$0.035 & 6.932$\pm$0.020 & 6.907$\pm$0.015 & 6.954$\pm$0.066$^\dagger$ \\
HIP~1481 & TH & 00182612-6328389 & F8/G0V & 4 & 7.460$\pm$0.010 & 0.537$\pm$0.008 &  & 2,2 & 6.462$\pm$0.018 & 6.248$\pm$0.036 & 6.149$\pm$0.017 & 6.141$\pm$0.048 & 6.102$\pm$0.023 & 6.138$\pm$0.015 & 5.746$\pm$0.031$^\dagger$ \\
TYC~1186-706-1 & BP & 00233468+2014282 & K7V(e) & 3 & 10.842$\pm$0.095 &  &  & 5 & 8.138$\pm$0.020 & 7.498$\pm$0.018 & 7.337$\pm$0.021 & 7.181$\pm$0.028 & 7.200$\pm$0.021 & 7.130$\pm$0.016 & 6.976$\pm$0.082 \\
HIP~1910 & TH & 00240899-6211042 & M0Ve & 6 & 11.330$\pm$0.015 & 1.390$\pm$0.015 & 1.840$\pm$0.020 & 2,2,2 & 8.385$\pm$0.026 & 7.708$\pm$0.034 & 7.494$\pm$0.021 & 7.354$\pm$0.026 & 7.306$\pm$0.019 & 7.226$\pm$0.016 & 7.110$\pm$0.087 \\
HIP~1993 & TH & 00251465-6130483 & M0Ve & 6 & 11.260$\pm$0.020 & 1.350$\pm$0.020 &  & 2,2 & 8.615$\pm$0.027 & 7.943$\pm$0.040 & 7.749$\pm$0.026 & 7.606$\pm$0.025 & 7.594$\pm$0.020 & 7.515$\pm$0.016 & 7.505$\pm$0.111 \\
HIP~2729 & TH & 00345120-6154583 & K5V & 4 & 9.560$\pm$0.010 & 1.050$\pm$0.020 & 1.380$\pm$0.010 & 2,2,6 & 7.337$\pm$0.018 & 6.721$\pm$0.034 & 6.533$\pm$0.018 & 6.427$\pm$0.044 & 6.443$\pm$0.019 & 6.405$\pm$0.015 & 6.279$\pm$0.048 \\
\enddata
\tablecomments{ Groups:
(BP) -- $\beta$~Pic Moving Group; 
(EC) -- $\eta$~Cha Cluster;
(TWA) -- TW Hydra Association;
(TH) -- Tucana--Horologium Association;
References for Spectral Type and optical $BVI_C$ photometry:
(1) \cite{houk1978}; 
(2) \cite{perryman1997};
(3) This Work
(4) \cite{houk1975};
(5) Converted from Tycho-2 using \cite{mamajek2002,mamajek2002e};
(6) \cite{torres2006};
(7) \cite{hawley1996};
(8) \cite{riaz2006};
(9) \cite{gray2006};
(10) \cite{jeffries1995};
(11) \cite{henden2012};
(12) \cite{zuckerman2004};
(13) \cite{weis1993};
(14) \cite{vyssotsky1956};
(15) \cite{robertson1987};
(16) \cite{houk1999};
(17) \cite{stephenson1969};
(18) \cite{gray1989};
(19) \cite{houk1988};
(20) \cite{schlieder2012b};
(21) \cite{rice2010b};
(22) \cite{luhman2004};
(23) \cite{lyo2004};
(24) \cite{lawson2001};
(25) \cite{lawson2002};
(26) \cite{stephenson1986};
(27) \cite{reid1995};
(28) \cite{koen2010};
(29) \cite{rodriguez2011};
(30) \cite{schneider2012b};
(31) \cite{scholz2005};
(32) \cite{teixeira2008};
(33) \cite{looper2010a};
(34) \cite{white2004};
(35) \cite{barrado2006};
(36) \cite{shkolnik2011};
(37) \cite{kastner2008};
(38) \cite{looper2007};
(39) \cite{song2002}; \\
($^a$) $JHK_S$ photometry from the 2MASS All-Sky Point Source Catalog
\citep{cutri2003, skrutskie2006};  \\
($^b$) $W1\,W2\,W3\,W4$ photometry from {\it WISE} All Sky Data Release \citep{cutri2012}; \\
($^\dagger$) Photometry excluded due to identified infrared excess 
\citep{hutchinson1990, megeath2005, riaz2006, sicilia-aguilar2009, 
       rebull2008, gautier2008, zuckerman2011, schneider2012a, schneider2012b}.
The binary pair HIP~10679 and HIP~10680 were studied in
\cite{rebull2008}, with only HIP~10679 identified as having a
24~$\mu$m excess.  However, they are separated by $\sim$14$\arcsec$
and thus it is likely that the HIP~10680 $W4$ photometry is
contaminated by the HIP~10679 $W4$ excess, so we exclude the HIP~10680
$W4$ photometry as well.
}
\end{deluxetable}
\clearpage
\end{landscape}

\clearpage
\begin{deluxetable}{llll}
\tabletypesize{\scriptsize}
\tablewidth{0pt}
\tablecaption{Spectral standard stars used for classification.\label{tbl:spstands}}
\tablehead{
  \colhead{Standard} &
  \colhead{Spectral} &
  \colhead{Telescope/Source} & 
  \colhead{References} \\
  \colhead{} & 
  \colhead{Type} & 
  \colhead{} & 
  \colhead{}
}
\startdata
  HD 8512           & K0IIIb & SMARTS 1.5m/Stony Brook & 1 \\
  HD 3651           & K0V    & SMARTS 1.5m/Stony Brook & 2 \\
  HD 10476          & K1V    & SMARTS 1.5m/Stony Brook & 1 \\
  HD 153210         & K2III  & SMARTS 1.5m/Stony Brook & 1 \\
  HD 22049          & K2V    & SMARTS 1.5m/Stony Brook & 1 \\
  HD 16160          & K3V    & SMARTS 1.5m/Stony Brook & 1 \\
  $\alpha$ Sct      & K3III  & SMARTS 1.5m/Rochester   & 1 \\
  TW PsA            & K4Ve   & SMARTS 1.5m/Rochester   & 1 \\
\enddata
\tablecomments{
References: (1) \cite{keenan1989}; (2) \cite{gray2003};
(3) \cite{gray2006}; (4) \cite{henry2002} \\
Spectral standards for F- and G-type stars were taken from 
Table~2 of \cite{pecaut2012}.
}
\end{deluxetable}

\begin{deluxetable}{lllll}
\tabletypesize{\scriptsize}
\tablewidth{0pt}
\tablecaption{Observations and New Spectral Types\label{tbl:sptypes}}
\tablehead{
  \colhead{Object} & 
  \colhead{Spectral Type} & 
  \colhead{Spectral} &
  \colhead{Spectral Type} & 
  \colhead{Ref.} \\
  \colhead{} & 
  \colhead{(This Work)} &
  \colhead{Coverage} & 
  \colhead{(Literature)} &
  \colhead{}
}
\startdata
HR 9              & F3V       & Blue      & F3Vn        & 1 \\
TYC 1186-706-1    & K7V(e)    & Blue      & K5          & 2  \\
HIP 10679         & G3V       & Blue	  & G2V         & 3 \\
HIP 10680         & F7V       & Blue	  & F7V         & 3 \\
HIP 12545         & K5IVe     & Blue/Red  & K6Ve        & 4  \\
HIP 12925         & F7V       & Blue      & F8          & 5 \\
GJ 3305           & M0Ve      & Blue/Red  & M1.1        & 6  \\
V1005 Ori         & K8IVe     & Blue/Red  & M0Ve        & 4  \\
\enddata
\tablecomments{
References: 
(1) \cite{gray2006}; 
(2) \cite{stephenson1986};  
(3) \cite{harlan1969}; 
(4) \cite{torres2006}; 
(5) \cite{cannon1918a}; 
(6) \cite{shkolnik2009}; 
(7) \cite{luhman2004}; 
(8) \cite{sterzik1999}; 
(9) \cite{white2004}; 
(10) \cite{reid2003}; \\
(11) \cite{zuckerman2001}; 
(12) \cite{strassmeier2000}; 
(13) \cite{stephenson1977}; 
(14) \cite{lepine2009}; 
(15) \cite{keenan1989};
$^a$ Our blue spectrum of this star was classified as G5V while our red 
spectrum was classified as a K1V.  Therefore we adopted an overall spectral type 
of G5V,K1V. \\
$^b$ Our spectrum of this star did not cover the Na doublet feature; \\
$^c$ \cite{mccarthy2012} were unable to detect Li in a high-resolution spectrum
of this star, casting doubt on its membership in $\beta$~Pic; however, we 
retain it as a member for the purposes of this study.
}
\end{deluxetable}

\begin{deluxetable}{lrrrrrrrrrrrrr}
\tabletypesize{\scriptsize}
\tablewidth{0pt}
\setlength{\tabcolsep}{0.02in}
\tablecaption{Intrinsic Colors of O9-M9 Dwarfs and Adopted \teff, Bolometric Correction Values\label{tbl:ic_teff}}
\tablehead{
 \colhead{SpT} &
 \colhead{\teff} &
 \colhead{BC$_V$} &
 \colhead{$U$--$B$} &
 \colhead{$B$--$V$} &
 \colhead{$V$--$R_C$} &
 \colhead{$V$--$I_C$} &
 \colhead{$V$--$J$} &
 \colhead{$V$--$H$} &
 \colhead{$V$--$K_S$} &
 \colhead{$K$--$W1$} &
 \colhead{$K$--$W2$} &
 \colhead{$K$--$W3$} &
 \colhead{$K$--$W4$} \\
 \colhead{} &
 \colhead{(K)} &
 \colhead{(mag)} &
 \colhead{(mag)} &
 \colhead{(mag)} &
 \colhead{(mag)} &
 \colhead{(mag)} &
 \colhead{(mag)} &
 \colhead{(mag)} &
 \colhead{(mag)} &
 \colhead{(mag)} &
 \colhead{(mag)} &
 \colhead{(mag)} &
 \colhead{(mag)}
}
\startdata
O9V   & 34000 & -3.20 & -1.114 & -0.318 &  ...   & -0.369 & -0.765 & -0.929 & -1.000 &  ...  &  ...  &  ...   &  ...  \\
O9.5V & 32000 & -3.06 & -1.087 & -0.312 &  ...   & -0.361 & -0.747 & -0.908 & -0.977 &  ...  &  ...  &  ...   &  ...  \\
B0V   & 31500 & -3.03 & -1.067 & -0.307 &  ...   & -0.355 & -0.732 & -0.891 & -0.958 &  ...  &  ...  &  ...   &  ...  \\
B0.5V & 29000 & -2.87 & -1.026 & -0.295 &  ...   & -0.338 & -0.697 & -0.850 & -0.913 &  ...  &  ...  &  ...   &  ...  \\
B1V   & 26000 & -2.61 & -0.995 & -0.278 & -0.115 & -0.325 & -0.667 & -0.815 & -0.874 &  ...  &  ...  &  ...   &  ...  \\
B1.5V & 24800 & -2.43 & -0.910 & -0.252 & -0.114 & -0.281 & -0.573 & -0.705 & -0.752 &  ...  &  ...  &  ...   &  ...  \\
B2V   & 20600 & -2.06 & -0.790 & -0.210 & -0.094 & -0.230 & -0.457 & -0.570 & -0.602 &  ...  &  ...  &  ...   &  ...  \\
B2.5V & 18500 & -1.79 & -0.732 & -0.198 & -0.087 & -0.210 & -0.413 & -0.518 & -0.544 &  ...  &  ...  &  ...   &  ...  \\
\enddata
\end{deluxetable}

\clearpage
\begin{landscape}
\begin{deluxetable}{rllllllllllllllll}
\tabletypesize{\scriptsize}
\setlength{\tabcolsep}{0.02in}
\tablewidth{0pt}
\tablecaption{Synthetic Color Indices From BT-Settl and ATLAS9 models\label{tbl:syncolors}}
\tablehead{
  \colhead{\teff} &
  \colhead{$\log(g)$} &
  \colhead{$U$--$B$} &
  \colhead{$B$--$V$} &
  \colhead{$V$--$I_C$} &
  \colhead{$R_C$--$I_C$} &
  \colhead{$J$--$H$} &
  \colhead{$H$--$K_S$} &
  \colhead{$V$--$K_S$} &
  \colhead{$K$--$W1$} &
  \colhead{$K$--$W2$} &
  \colhead{$K$--$W3$} &
  \colhead{$K$--$W4$} &
  \colhead{$V$--$V_T$} &
  \colhead{$B_T$--$V_T$} &
  \colhead{$V$--$H_{\rm p}$} &
  \colhead{Model} \\
  \colhead{(K)} &
  \colhead{(dex)} & 
  \colhead{(mag)} & 
  \colhead{(mag)} & 
  \colhead{(mag)} & 
  \colhead{(mag)} & 
  \colhead{(mag)} & 
  \colhead{(mag)} & 
  \colhead{(mag)} & 
  \colhead{(mag)} & 
  \colhead{(mag)} & 
  \colhead{(mag)} & 
  \colhead{(mag)} & 
  \colhead{(mag)} & 
  \colhead{(mag)} & 
  \colhead{(mag)} &
  \colhead{} 
}
\startdata
1400 & 3.5 & 3.475 &  3.191 & 5.213 & 2.454 & 1.916 & 1.344 & 13.151 & 1.379 & 1.967 & 2.652 & 2.921 & -0.405 & 3.747 & 0.979 & BT-Settl \\
1400 & 4.0 & 4.909 &  0.795 & 7.250 & 2.854 & 1.149 & 0.635 & 13.788 & 0.983 & 1.083 & 2.226 & 2.292 &  0.091 & 1.668 & 2.550 & BT-Settl \\
1400 & 4.5 & 4.959 & -0.165 & 7.918 & 3.184 & 0.919 & 0.306 & 13.929 & 0.923 & 1.148 & 2.326 & 2.426 &  0.561 & 1.123 & 2.903 & BT-Settl \\
1400 & 5.0 & 4.957 & -0.945 & 8.421 & 3.433 & 0.893 & 0.098 & 14.184 & 0.873 & 1.363 & 2.474 & 2.631 &  1.085 & 0.836 & 3.189 & BT-Settl \\
\enddata
\tablecomments{
  All synthetic colors are computed using solar metallicity models. The
  BT-Settl model colors presented here adopt the \cite{asplund2009} solar composition, 
  whereas the ATLAS9 model colors presented here adopt the \cite{grevesse1998} solar composition. 
  Table~\ref{tbl:syncolors} is published in its entirety in the electronic edition of ApJS.
  A portion is shown here for guidance regarding its form and content.
}
\end{deluxetable}
\clearpage
\end{landscape}

\begin{deluxetable}{lrrrrrrrrrrrr}
\tabletypesize{\scriptsize}
\setlength{\tabcolsep}{0.02in}
\tablewidth{0pt}
\tablecaption{Intrinsic colors of 5-30 Myr old Stars and Adopted T$_{\rm eff}$, Bolometric Correction Values\label{tbl:young_colors}}
\tablehead{
  \colhead{Spec.} & 
  \colhead{T$_{\rm eff}$} &
  \colhead{$B$--$V$} & 
  \colhead{$V$--$I_C$} & 
  \colhead{$V$--$K_S$} & 
  \colhead{$J$--$H$} & 
  \colhead{$H$--$K_S$} &
  \colhead{$K_S$--$W1$} &
  \colhead{$K_S$--$W2$} &
  \colhead{$K_S$--$W3$} &
  \colhead{$K_S$--$W4$} &
  \colhead{BC$_V$} & 
  \colhead{BC$_J$} \\
  \colhead{Type} & 
  \colhead{(K)} & 
  \colhead{(mag)} & 
  \colhead{(mag)} & 
  \colhead{(mag)} & 
  \colhead{(mag)} & 
  \colhead{(mag)} & 
  \colhead{(mag)} & 
  \colhead{(mag)} & 
  \colhead{(mag)} & 
  \colhead{(mag)} &
  \colhead{(mag)} &
  \colhead{(mag)} 
}
\startdata
F0 & 7280 & 0.28 & 0.34 & 0.73 & 0.11 & 0.06 & 0.04 & 0.03 & 0.00 & 0.08 & 0.01 & 0.57  \\
F1 & 6990 & 0.34 & 0.39 & 0.89 & 0.14 & 0.07 & 0.04 & 0.04 & 0.00 & 0.09 & 0.00 & 0.68  \\
F2 & 6710 & 0.38 & 0.43 & 0.99 & 0.15 & 0.08 & 0.04 & 0.05 & 0.00 & 0.09 & -0.01 & 0.75  \\
F3 & 6660 & 0.41 & 0.45 & 1.01 & 0.16 & 0.08 & 0.04 & 0.05 & 0.01 & 0.09 & -0.01 & 0.76  \\
F4 & 6590 & 0.43 & 0.48 & 1.05 & 0.17 & 0.08 & 0.04 & 0.06 & 0.01 & 0.09 & -0.01 & 0.79  \\
F5 & 6420 & 0.47 & 0.51 & 1.14 & 0.19 & 0.08 & 0.04 & 0.03 & 0.01 & 0.10 & -0.02 & 0.85  \\
F6 & 6250 & 0.50 & 0.55 & 1.25 & 0.21 & 0.09 & 0.04 & 0.04 & 0.02 & 0.10 & -0.04 & 0.91  \\
F7 & 6140 & 0.53 & 0.58 & 1.31 & 0.22 & 0.09 & 0.05 & 0.04 & 0.02 & 0.10 & -0.05 & 0.95  \\
\enddata
\end{deluxetable}

\begin{deluxetable}{llllll}
\tabletypesize{\scriptsize}
\tablewidth{0pt}
\tablecaption{\teff\, Comparison: SEDF versus Diameter-Derived \teff\label{tbl:ad_teffs}}
\tablehead{
  \colhead{Object} & 
  \colhead{SpT} & 
  \colhead{Ref.} & 
  \colhead{\teff$^a$} &
  \colhead{\teff$^b$} &
  \colhead{m$_{bol}^c$}  \\
  \colhead{} & 
  \colhead{} & 
  \colhead{} &
  \colhead{(K)} & 
  \colhead{(K)} &
  \colhead{(mag)}
}
\startdata
GJ~15A   & M1.5V    & 1 & 3535$\pm$14 & 3567$\pm$11 & 6.560$\pm$0.014  \\
GJ~33    & K2.5V    & 2 & 5047$\pm$19 & 4950$\pm$14 & 5.459$\pm$0.017  \\
GJ~53A   & K1V Fe-2 & 3 & 5362$\pm$22 & 5348$\pm$26 & 4.981$\pm$0.016  \\
GJ~75    & G9V      & 3 & 5395$\pm$20 & 5398$\pm$75 & 5.452$\pm$0.018  \\
GJ~105   & K3V      & 2 & 4827$\pm$18 & 4662$\pm$17 & 5.406$\pm$0.015  \\
GJ~144   & K2V(k)   & 2 & 5097$\pm$19 & 5077$\pm$35 & 3.457$\pm$0.016  \\
GJ~166A  & K0.5V    & 2 & 5162$\pm$21 & 5147$\pm$14 & 4.169$\pm$0.016  \\
GJ~205   & M1.5V    & 1 & 3703$\pm$36 & 3801$\pm$9  & 6.443$\pm$0.016  \\
\enddata
\tablecomments{
$^a$: \teff\, from this work using the SEDF method.  See Section~\ref{sec:sedf} for details. 
$^b$: \teff\, from \cite{boyajian2012b} computed using direct angular diameter measurements.
$^c$: apparent bolometric magnitude estimated from our SED fit.\\
Spectral Type References: (1) \citet{henry2002}; (2) \citet{gray2006}; (3) \citet{gray2003}; 
\citet{keenan1989}, (5) \cite{torres2006};
}
\end{deluxetable}

\begin{deluxetable}{lll}
\tabletypesize{\scriptsize}
\tablewidth{0pt}
\tablecaption{Objects Rejected From SED-\teff\, fitting\label{tbl:sedrejects}}
\tablehead{
  \colhead{Object} &
  \colhead{Rejection Reason} 
}
\startdata
HD~139084B             & Uncertain photometry resulting in poorly constrained \teff \\
HD~164249B             & Uncertain photometry resulting in poorly constrained \teff \\
RECX~11                & $K_S$ band excess \\
RECX~15                & $K_S$ band excess \\
RECX~16                & {\it IRAC} 3.6$\mu$m and 4.5$\mu$m excess \\
TWA~27                 & {\it IRAC} 3.6$\mu$m and 4.5$\mu$m excess \\
TWA~29                 & Only three good bands of photometry ($JHK_S$) \\
TWA~30A                & time-variable extinction \\
TWA~30B                & time-variable NIR excess \\
TWA~31                 & Only three good bands of photometry ($JHK_S$) \\
TWA~34                 & Only three good bands of photometry ($JHK_S$)
\enddata
\end{deluxetable}

\clearpage
\begin{landscape}
\begin{deluxetable}{lllrrrrrlrrrrr}
\tabletypesize{\tiny}
\setlength{\tabcolsep}{0.02in}
\tablewidth{0pt}
\tablecaption{\teff, Bolometric Magnitudes, Bolometric Corrections and Angular Diameter Estimates From SED Fitting\label{tbl:teff_fit}}
\tablehead{
  \colhead{Object} &
  \colhead{2MASS} &
  \multicolumn{6}{c}{BT-Settl} &
  \multicolumn{6}{c}{Kurucz} \\
  \cline{3-8}
  \cline{9-14}
  \colhead{} &
  \colhead{} &
  \colhead{\teff} &
  \colhead{$\theta$} &
  \colhead{$BC_V$} &
  \colhead{$BC_J$} &
  \colhead{$m_{bol}$} &
  \colhead{$\log(L/L_{\odot})$} &
  \colhead{\teff} &
  \colhead{$\theta$} &
  \colhead{$BC_V$} &
  \colhead{$BC_J$} &
  \colhead{$m_{bol}$} &
  \colhead{$\log(L/L_{\odot})$} \\
  \colhead{} &
  \colhead{} &
  \colhead{(K)} &
  \colhead{($\mu$as)} &
  \colhead{(mag)} &
  \colhead{(mag)} &
  \colhead{(mag)} &
  \colhead{(dex)} &
  \colhead{(K)} &
  \colhead{($\mu$as)} &
  \colhead{(mag)} &
  \colhead{(mag)} &
  \colhead{(mag)} &
  \colhead{(dex)} 
}
\startdata
HIP~490 & 00055255-4145109 & 5990$\pm$16 & 250$\pm$1 & -0.06$\pm$0.02 & 0.98$\pm$0.02 & 7.45$\pm$0.02 & 0.114$\pm$0.048 & 5968$\pm$15 & 250$\pm$1 & -0.05$\pm$0.02 & 1.00$\pm$0.02 & 7.46$\pm$0.02 & 0.108$\pm$0.047  \\
HR~9 & 00065008-2306271 & 6796$\pm$25 & 346$\pm$1 & 0.00$\pm$0.02 & 0.74$\pm$0.03 & 6.19$\pm$0.02 & 0.617$\pm$0.033 & 6773$\pm$21 & 348$\pm$1 & 0.01$\pm$0.02 & 0.75$\pm$0.03 & 6.20$\pm$0.02 & 0.614$\pm$0.032  \\
HIP~1113 & 00135300-7441178 & 5434$\pm$15 & 180$\pm$1 & -0.17$\pm$0.02 & 1.18$\pm$0.03 & 8.59$\pm$0.02 & -0.238$\pm$0.074 & 5401$\pm$16 & 180$\pm$1 & -0.15$\pm$0.02 & 1.21$\pm$0.03 & 8.61$\pm$0.02 & -0.247$\pm$0.074  \\
HIP~1481 & 00182612-6328389 & 6150$\pm$18 & 242$\pm$1 & -0.06$\pm$0.02 & 0.94$\pm$0.02 & 7.41$\pm$0.02 & 0.177$\pm$0.045 & 6123$\pm$17 & 242$\pm$1 & -0.04$\pm$0.02 & 0.96$\pm$0.02 & 7.42$\pm$0.02 & 0.171$\pm$0.045  \\
TYC~1186-706-1 & 00233468+2014282 & 4055$\pm$15 & 189$\pm$1 & -1.09$\pm$0.10 & 1.61$\pm$0.03 & 9.75$\pm$0.03 & ... & 3888$\pm$42 & 197$\pm$1 & -1.00$\pm$0.11 & 1.70$\pm$0.05 & 9.84$\pm$0.05 & ...  \\
HIP~1910 & 00240899-6211042 & 3823$\pm$18 & 188$\pm$2 & -1.31$\pm$0.02 & 1.64$\pm$0.03 & 10.02$\pm$0.03 & -0.659$\pm$0.288 & 3805$\pm$24 & 189$\pm$2 & -1.31$\pm$0.02 & 1.64$\pm$0.03 & 10.03$\pm$0.02 & -0.660$\pm$0.288  \\
HIP~1993 & 00251465-6130483 & 4015$\pm$14 & 159$\pm$1 & -1.09$\pm$0.03 & 1.56$\pm$0.03 & 10.17$\pm$0.03 & -0.845$\pm$0.222 & 3975$\pm$17 & 161$\pm$1 & -1.07$\pm$0.03 & 1.57$\pm$0.03 & 10.19$\pm$0.03 & -0.850$\pm$0.222  \\
HIP~2729 & 00345120-6154583 & 4376$\pm$10 & 252$\pm$1 & -0.76$\pm$0.02 & 1.46$\pm$0.02 & 8.80$\pm$0.02 & -0.330$\pm$0.087 & 4318$\pm$9 & 256$\pm$1 & -0.74$\pm$0.01 & 1.48$\pm$0.02 & 8.82$\pm$0.02 & -0.339$\pm$0.087  \\
\enddata
\tablecomments{
    \teff\, values were fit at $\log(g)=4.3$.  \\
    $a$ Indicates $\log(L/L_{\odot})$ estimates use parallaxes from \cite{weinberger2012}.\\
    $b$ Indicates $\log(L/L_{\odot})$ estimate uses parallax from \cite{teixeira2008}.\\
    $c$ Indicates  $\log(L/L_{\odot})$ estimate uses parallax from \cite{teixeira2009}.\\
    In addition, we have used a weighted mean distance of 94.3$\pm$1.2~pc for $\eta$~Cha 
    cluster members.  For HD~139084B, HD~164249B, AZ~Cap, and HD~222259B
    we have adopted parallaxes from their brighter companions.
}
\end{deluxetable}
\clearpage
\end{landscape}

\clearpage
\begin{landscape}
\begin{deluxetable}{lllllllllll}
\tabletypesize{\tiny}
\setlength{\tabcolsep}{0.02in}
\tablewidth{0pt}
\tablecaption{\teff, Bolometric Correction, and Bolometric Magnitude Polynomial Coefficients\label{tbl:teffbcpoly} for 5-30 Myr Old Stars}
\tablehead{
  \colhead{$Y$} &
  \colhead{$X$} &
  \colhead{Range} &
  \colhead{$a_0$} &
  \colhead{$a_1$} &
  \colhead{$a_2$} &
  \colhead{$a_3$} &
  \colhead{$a_4$} &
  \colhead{$a_5$} &
  \colhead{$a_6$} &
  \colhead{$a_7$}
}
\startdata
\teff  & $V$--$K_S$ & $1.0 < V-K_S < 6.7$ &  9.323430$\times10^{3}$   &  -3.516011$\times10^{3}$  &  1.046787$\times10^{3}$  & -1.863349$\times10^{2}$  &  1.641182$\times10^{1}$   & -5.188853$\times10^{-1}$  & ... & ... \\
\teff  & $V$--$J$   & $0.8 < V-J < 5.8$   &  9.593475$\times10^{3}$   &  -5.095204$\times10^{3}$  &  2.053259$\times10^{3}$  & -4.813940$\times10^{2}$  &  5.816754$\times10^{1}$   & -2.779565$\times10^{0}$  & ... & ... \\
$BC_V$ & $V$--$K_S$ & $1.0 < V-K_S < 6.7$ & -7.443324$\times10^{-2}$  &  2.471780$\times10^{-1}$ &  -1.923234$\times10^{-1}$  & 1.318867$\times10^{-2}$ &  -3.630511$\times10^{-4}$ & ...  & ... & ... \\
$BC_V$ & \teff      & $2750 <$ \teff $< 7350$ &  -2.855844$\times10^{2}$  &  3.832453$\times10^{-1}$ & -2.225832$\times10^{-4}$ &  7.150667$\times10^{-8}$ & -1.364193$\times10^{-11}$  & 1.542389$\times10^{-15}$ &  -9.566224$\times10^{-20}$ & 2.511807$\times10^{-24}$ \\
$BC_J$ & $V$--$K_S$ & $1.0 < V-K_S < 6.7$ &  -4.805196$\times10^{-1}$  &  1.842350$\times10^{0}$ & -7.837156$\times10^{-1}$  & 1.859281$\times10^{-1}$ & -2.153500$\times10^{-2}$ & 9.489583$\times10^{-4}$  & ...  & ...  \\
$BC_J$ & $V$--$J$   & $0.8 < V-J < 5.8$ &  -4.557821$\times10^{-1}$  &  2.299875$\times10^{0}$ & -1.191653$\times10^{0}$ & 3.442879$\times10^{-1}$ & -4.932544$\times10^{-2}$ &  2.724400$\times10^{-3}$  & ... & ... \\
\enddata
\tablecomments{
 $Y = a_0 + a_1X + a_2X^2 + a_3X^3 + a_4X^4 + a_4X^4 + a_5X^5 + a_6X^6 + a_7X^7$ 
}
\end{deluxetable}
\clearpage
\end{landscape}

\end{document}